\PassOptionsToPackage{table,xcdraw}{xcolor}  
\documentclass[sigconf]{acmart}
\AtBeginDocument{%
  }

\setcopyright{acmlicensed}
\copyrightyear{2018}
\acmYear{2018}
\acmDOI{XXXXXXX.XXXXXXX}
\acmConference[arXiv '25]{}{}{}
\acmISBN{978-1-4503-XXXX-X/2025/10}

\renewcommand\footnotetextcopyrightpermission[1]{}

\usepackage{amsmath, amssymb, amsfonts, bm}

\usepackage{booktabs, multirow, graphicx, float, enumitem}

\usepackage[font=small, skip=5pt]{caption}
\usepackage{subcaption}  

\usepackage{tipa}

\begin{document}

\title{EchoMask: Speech-Queried Attention-based Mask Modeling for Holistic Co-Speech Motion Generation}

\author{
Xiangyue Zhang*$^{1,2}$ \quad Jianfang Li*$^2$ \quad Jiaxu Zhang$^{1}$ \\ \quad Jianqiang Ren$^{2}$ \quad Liefeng Bo$^{2}$ \quad Zhigang Tu$^{1\dagger}$ \\
$^1$Wuhan University \quad $^2$ Tongyi Lab, Alibaba Group \quad
\vspace{0.5em} \\
\textbf{Project page:} \href{https://xiangyuezhang.com/EchoMask/}{https://xiangyuezhang.com/EchoMask/}
}
\renewcommand{\shortauthors}{Zhang et al.}

\begin{abstract}
Masked modeling has shown promise in co-speech gesture generation. However, it struggles to identify semantically significant frames for effective motion masking.
In this work, we propose a speech-queried attention-based mask modeling framework for holistic co-speech gesture generation. Our key insight is to leverage motion-aligned speech features to guide the masked motion modeling process, selectively masking rhythm-related and semantically expressive motion frames.
Specifically, we first propose a motion-audio alignment module (MAM) to construct a latent motion-audio joint space. In this space, both low-level and high-level speech features are projected, enabling motion-aligned speech representation using learnable speech queries. Then, a speech-queried attention mechanism (SQA) is introduced to compute frame-level attention scores through interactions between motion keys and speech queries, guiding selective masking toward motion frames with high attention scores.
Finally, the motion-aligned speech features are also injected into the generation network to facilitate co-speech motion generation. Qualitative and quantitative evaluations confirm that our method outperforms existing state-of-the-art approaches, successfully producing high-quality co-speech motion.
\end{abstract}


\begin{teaserfigure}
  \includegraphics[width=\textwidth]{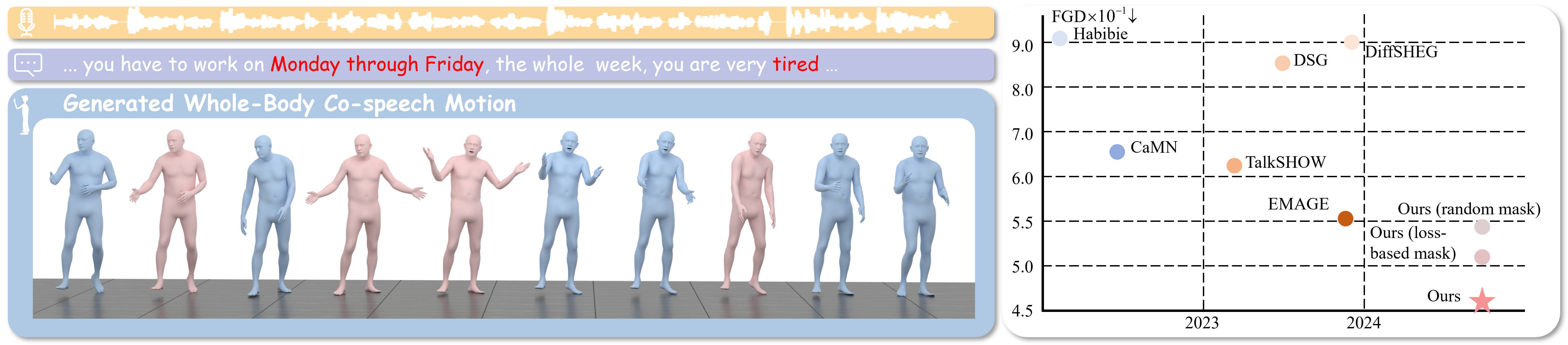}
  \caption{\textbf{On the left}, our EchoMask generates expressive, semantically aligned whole-body co-speech motions from speech input. \textbf{On the right}, our method outperforms previous approaches by leveraging speech-queried attention mask modeling and motion-audio embedding, surpassing both random and loss-based strategies.}
  \Description{Enjoying the baseball game from the third-base
  seats. Ichiro Suzuki preparing to bat.}
  \label{fig:teaser}
\end{teaserfigure}


\maketitle

\section{Introduction}
Holistic co-speech motion generation, integrating non-verbal cues such as body poses \cite{chhatre2024emotional,ginosar2019learning}, hand gestures \cite{liu2022learning, li2021audio2gestures}, and facial expressions \cite{danvevcek2023emotional,xing2023codetalker} aligned with speech, is an increasingly prominent field in computer vision. It holds immense value for various applications, including virtual avatars, gaming, immersive virtual live streaming, and robotics.

Generating detailed and expressive whole-body motions remains a computationally intensive and technically challenging problem. Masked motion modeling (MMM) \cite{guo2024momask,pinyoanuntapong2024bamm,pinyoanuntapong2024mmm} has recently gained attention as a promising framework to address these challenges \cite{liu2024emage,liu2024towards,jeonghgm3}. MMM begins by discretizing continuous motion sequences into compact motion tokens via vector quantization (VQ) \cite{van2017neural}, enabling efficient learning in a discrete latent space. During training, a subset of motion frames or tokens is masked, and the model is tasked with reconstructing the missing elements from the remaining context. A critical factor in the success of MMM lies in the masking strategy—specifically, which frames or tokens are selected for masking. Evidence from masked image modeling (MIM) \cite{bao2021beit,he2022masked,xie2022simmim} shows that mask patterns significantly affect model performance by shaping the focus of the learning process \cite{wang2023hard,xi2024global,kakogeorgiou2022hide}. However, current masking approaches in MMM often rely on random or loss-based guided strategies, which struggle to consistently identify semantically important motion segments. This limitation hampers the model's ability to generate expressive, temporally aligned motions, especially in co-speech scenarios where nuanced synchronization is essential.

By delving deeper into the underlying limitations, we identify the core bottleneck in advancing semantically grounded masked motion modeling: the effectiveness of semantically rich motion frames mask selection strategies. Existing methods \cite{guo2024momask,liu2024emage,liu2024towards,jeonghgm3} predominantly adopt random or loss-based masking strategies. Random masking often fails to target semantically meaningful regions due to the sparse distribution of semantic content in motion sequences. Loss-based masking, which prioritizes frames with high reconstruction error, assumes these are semantically rich. However, this assumption does not always hold—high reconstruction loss may simply reflect abrupt yet uninformative transitions. For instance, a sharp change in hand position might signal the end of a sentence rather than a meaningful gesture. As a result, such strategies struggle to accurately identify semantically significant frames, ultimately limiting the quality of speech-conditioned motion generation. Moreover, prior methods mask at the token level using discrete code indices, which lose fine-grained motion details and hinder accurate detection of frames critical for motion intent and speech alignment.

Based on this observation, we raise a central question: Can speech be used as a query to identify semantically important motion frames worth focusing on during masked modeling? To explore this, we propose a new masked motion modeling framework, EchoMask, for co-speech motion generation. Our core motivation is to leverage speech not only as a conditioning signal but also as an informative query that aligns with latent motion representations to pinpoint motion frames with high semantic relevance. 

In EchoMask, we introduce a speech-queried, attention-based masked motion modeling framework that leverages motion-aligned speech features to guide the masked motion modeling process, selectively masking semantically rich motion frames. To this end, we propose two key innovations: a hierarchical joint embedding module for motion-audio alignment (MAM) and a speech-queried attention mechanism (SQA) for keyframe selection.

Specifically, MAM is a hierarchical cross-modal alignment module, designed to project paired latent motion and audio inputs into a shared latent space to effectively align the heterogeneous motion and audio modalities. Inspired by CLIP-style dual-tower architectures and \cite{liu2024tango,li2024lamp}, MAM utilizes both low-level and high-level HuBERT features, interacting with learnable speech queries via self-attention and cross-attention to generate motion-aligned speech features. The speech queries and latent motion are passed through a shared transformer network, encouraging modality-invariant feature extraction. This design not only reduces parameter redundancy but also promotes more consistent alignment between speech and motion representations, facilitating improved generation fidelity.

\begin{figure}
    \centering
    \includegraphics[width=0.49\textwidth]{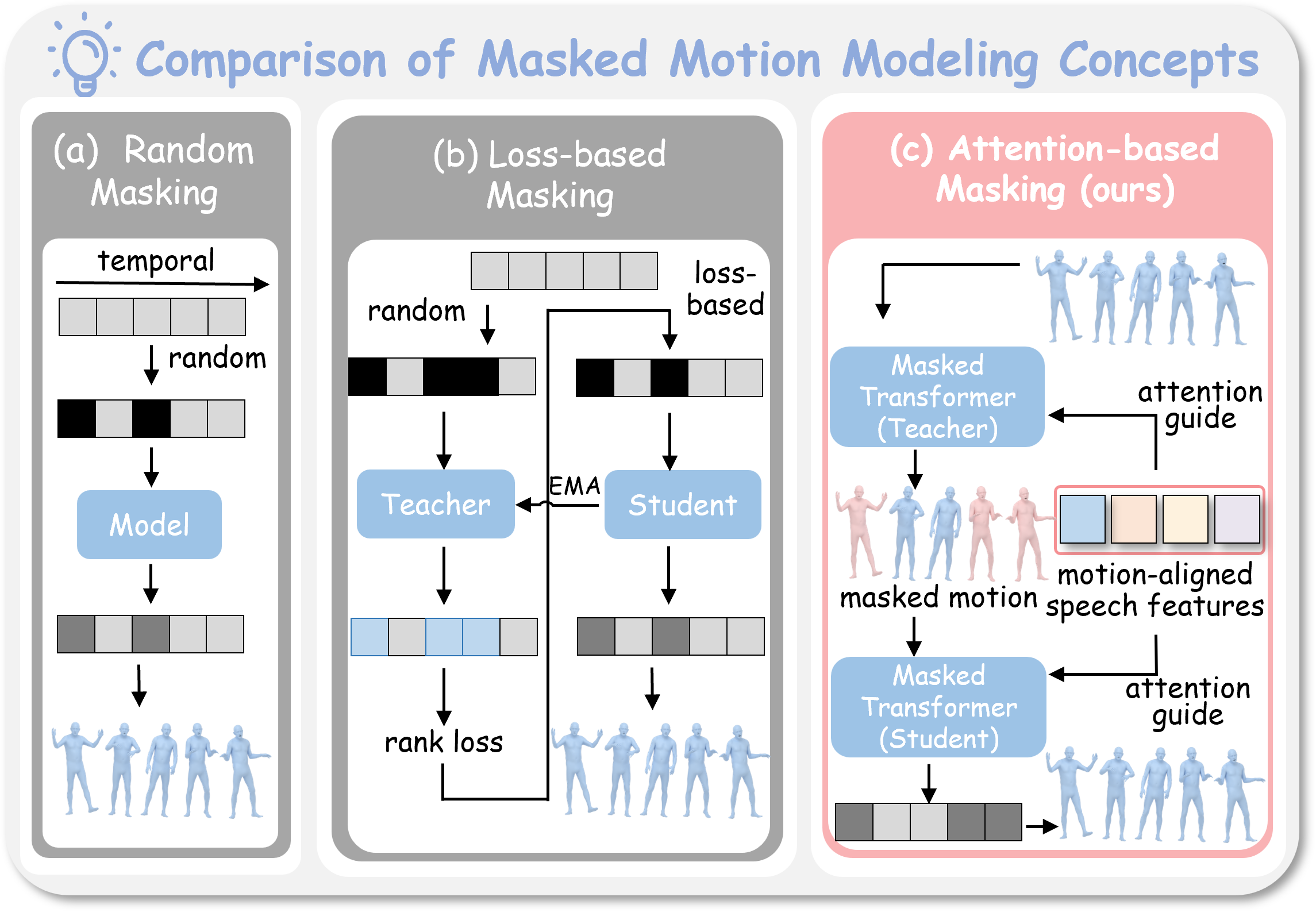}
    \caption{Comparison of masked motion modeling concepts, where three masking strategies are reported: (a) random masking, (b) loss-based masking, and (c) our proposed speech-queried attention-based masking. Unlike the previous approaches, our method selectively highlights motion frames (in red) that are more semantically aligned with the input speech.}
    \vspace{-15pt}
    \label{fig:previous}
\end{figure}

SQA then computes frame-level attention scores by modeling the interaction between motion keys and motion-aligned speech features produced by the MAM module. Frames receiving higher attention scores are identified as semantically informative and are preferentially masked. This targeted masking strategy encourages the model to focus on speech-relevant motion patterns during training, as illustrated in Figure~\ref{fig:previous}. By selectively masking keyframes based on attention scores, the model is guided to learn meaningful associations between speech and motion. Finally, the motion-aligned speech features are injected into the generation network, enhancing the quality and synchronization of the synthesized co-speech motion.

We comprehensively evaluate EchoMask on public co-speech motion generation benchmarks. Our qualitative and quantitative results demonstrate that EchoMask significantly outperforms existing methods in terms of semantic alignment, motion realism, and generation diversity.

Our contributions are summarized below:
\begin{itemize}[itemsep=0pt, parsep=0pt]
\sloppy  
\item We propose EchoMask for co-speech motion generation, a novel masked motion modeling framework that utilizes motion-aligned speech features to mask semantically important motion frames.
\item We introduce MAM, a hierarchical cross-modal alignment module that embeds motion and audio into a shared latent space, producing motion-aligned speech features. We also propose SQA, a speech-queried attention mechanism that computes frame-level attention scores, enabling selective identification of semantically important motion frames.
\item Extensive experiments demonstrate the superiority of EchoMask over state-of-the-art methods in terms of semantic alignment, motion quality, and generation diversity.
\end{itemize}
\section{Related Work}
\noindent\textbf{Holistic Co-speech Motion Generation}
Holistic co-speech motion generation involves producing coordinated movements of the face, hands, and torso from speech input. Most existing methods focus on isolated parts rather than generating fully integrated body motion. Early rule-based methods \cite{kipp2005gesture, kopp2006towards, cassell2001beat, huang2012robot} mapped speech to gestures using linguistic rules but required extensive manual effort. Recent data-driven approaches use deep generative models—such as GANs \cite{habibie2021learning, ojha2021few, ahuja2022low, wang2020minegan}, VQ-VAEs \cite{ao2022rhythmic, liu2022audio, shu2024eggesture}, normalizing flows \cite{ye2022audio, tan2024flowvqtalker}, and diffusion models \cite{zhu2023taming, alexanderson2023listen, yang2023diffusestylegesture, he2024co, chen2024enabling}—to learn complex motion distributions from data.

Habibie et al. \cite{habibie2021learning} introduced a CNN-based model that outputs 3D face, hand, and body motion, but coordination and diversity were limited. Later methods improved realism and synchronization using discrete latent codes. For instance, TalkSHOW \cite{yi2023generating} synchronizes body and hand gestures with a VQ-VAE, but handles facial expressions separately. DiffSHEG \cite{chen2024diffsheg} enables unidirectional information flow between face and body through separate encoders.

Masked modeling has further improved efficiency. EMAGE \cite{liu2024emage} uses random motion masking, while ProbTalk \cite{liu2024towards} leverages PQ-VAE \cite{wu2019learning} to reconstruct masked inputs. However, these methods rely on random or causal masking, which limits the model's ability to focus on semantically important motion frames. In our work, we propose a speech-queried attention-based mask modeling that utilizes motion-aligned speech features to identify semantically important motion frames for co-speech motion generation.

\noindent\textbf{Masking Strategies in Masked Modeling}
Masking plays a key role in masked modeling by influencing what the model learns. Existing strategies fall into three categories: (i) \textbf{Random masking}. Used in early models like BERT \cite{devlin2019bert} and extended to vision in ViT \cite{dosovitskiy2020image}, with refinements in BEiT \cite{bao2021beit}. Generative models such as MaskGIT \cite{chang2022maskgit} and Muse \cite{chang2023muse} also apply random masking for representation learning.  (ii) \textbf{Loss-based masking}. AdaMAE \cite{bandara2023adamae} assumes that semantically meaningful patches are harder to reconstruct and therefore prioritizes masking regions with high reconstruction loss to guide the model toward learning informative content. HPM \cite{wang2023hard} focuses on increasing task difficulty by selecting patches the model finds hard to reconstruct, using a masking strategy that progresses from easy to hard examples. (iii) \textbf{Attention-based masking}. AMT \cite{liu2023good}, and AttMask \cite{kakogeorgiou2022hide} use class tokens to mask semantically rich regions, while MILAN \cite{hou2022milan} applies CLIP-based knowledge distillation for high-importance patch selection. While effective, these methods struggle to flexibly identify semantically important content guided by external signals. Inspired by \cite{xi2024global}, we introduce speech-queried attention masking that computes frame-level attention scores by evaluating the interaction between motion keys and motion-aligned speech features to identify significant frames.

\section{Method}
\subsection{Preliminary on Masked Motion Modeling}
Masked motion modeling (MMM) draws inspiration from masked image modeling~\cite{wang2023hard, bandara2023adamae, chang2022maskgit, chang2023muse} and aims to learn expressive motion representations by reconstructing missing portions of a motion sequence based on visible frames or tokens. Given a motion sequence consisting of $N$ frames, we denote it as a set of frame-wise units $M = \{\mathbf{m}_i\}_{i=1}^{N}$, where each $\mathbf{m}_i$ represents either joint positions, rotation representations, or quantized latent tokens at frame $i$.

A masking ratio $r \in (0, 1)$ is applied to randomly select a subset of frames or tokens $\mathcal{S} \subset \{1, \dots, N\}$, such that $|\mathcal{S}| = \lfloor rN \rfloor$. The selected elements are masked and replaced with learnable mask frames or tokens. The remaining unmasked subset $S_r = \{\mathbf{m}_i : i \notin \mathcal{S}\}$ is passed through an encoder to extract contextual motion features. The model is then trained to reconstruct the masked content using both the encoded visible frames and the learnable mask representations. A decoder receives the encoded visible context and the mask embeddings to generate predictions for the masked frames or tokens. The training objective is to minimize the reconstruction error between the predicted and ground-truth motion values for the masked subset:
\begin{equation}
\mathcal{L}_{\mathrm{MMM}} = \sum_{i \in \mathcal{S}} \left\lVert \mathbf{m}_i^d - \mathbf{m}_i \right\rVert_2^2.
\end{equation}

where $\mathbf{m}_i^d$ denotes the reconstructed output of the decoder for the $i$-th masked frame or token. Through this process, the model is encouraged to learn the underlying spatiotemporal structure and semantic consistency of human motion. MMM thus provides a powerful pretext task for learning generalizable and context-aware motion representations.
\begin{figure}
    \centering
    \includegraphics[width=0.49\textwidth]{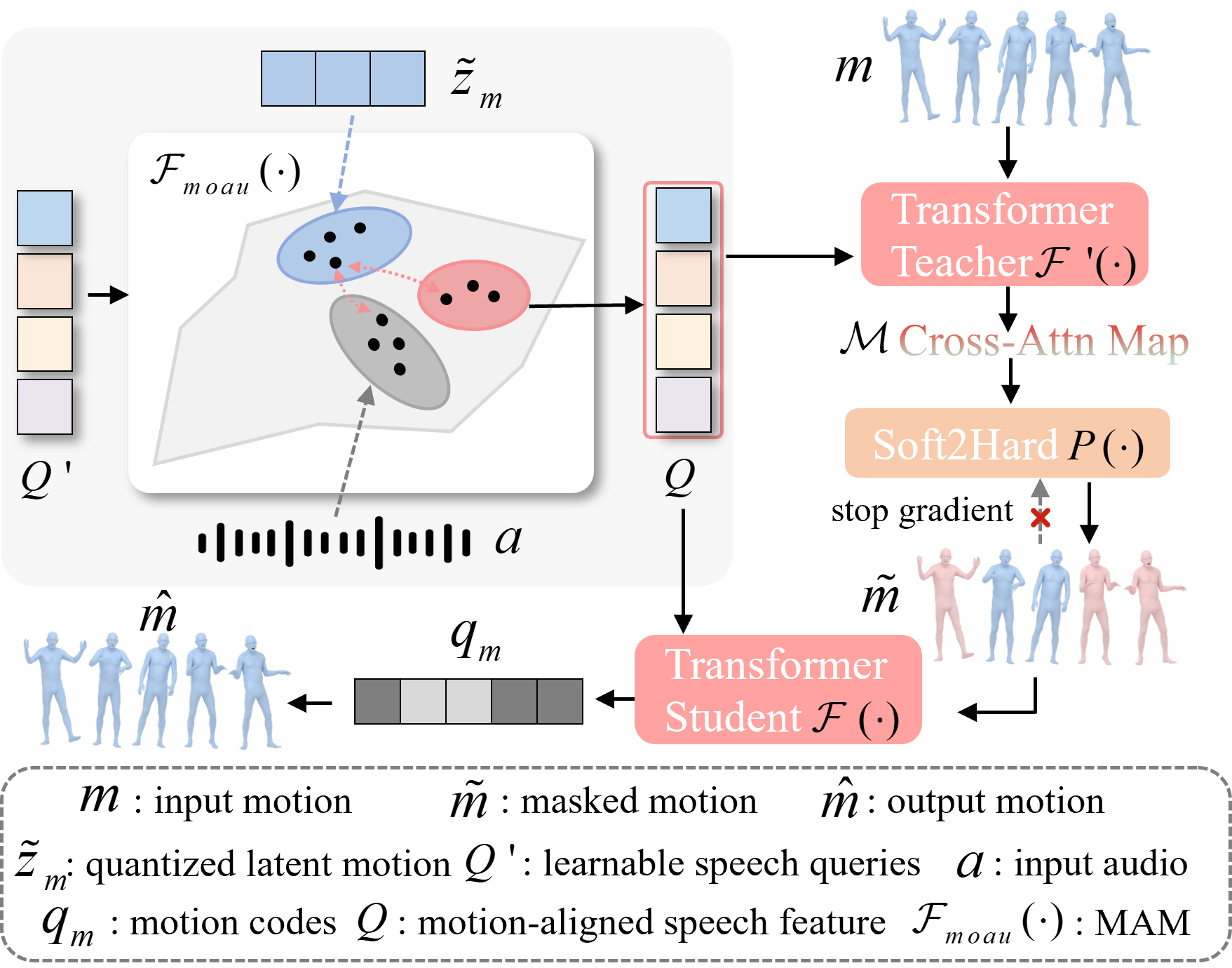}
    \caption{An overview of the EchoMask pipeline. Given audio \( a \) and quantized latent motion \( \tilde{z}_m \), MAM \( \mathcal{F}_{\text{moau}} \) produces motion-aligned speech features \( Q \) in a shared latent space. The Transformer Teacher \( \mathcal{F}' \) computes a cross-attention map \(\mathcal{M}\) between \( Q \) and input motion \(m\), which guides the Soft2Hard module \( P(\cdot) \) to mask semantically important frames. The Transformer Student \( \mathcal{F} \) then generates motion codes \(q_m\) from the masked motion \(\tilde{m}\) and \( Q \), and the output motion \(\hat{m}\) is decoded via the RVQ-VAE decoder.}
    \vspace{-15pt}
    \label{fig:overview}
\end{figure}

\subsection{Overview}
\begin{figure*}
    \centering
    \includegraphics[width=0.98\textwidth]{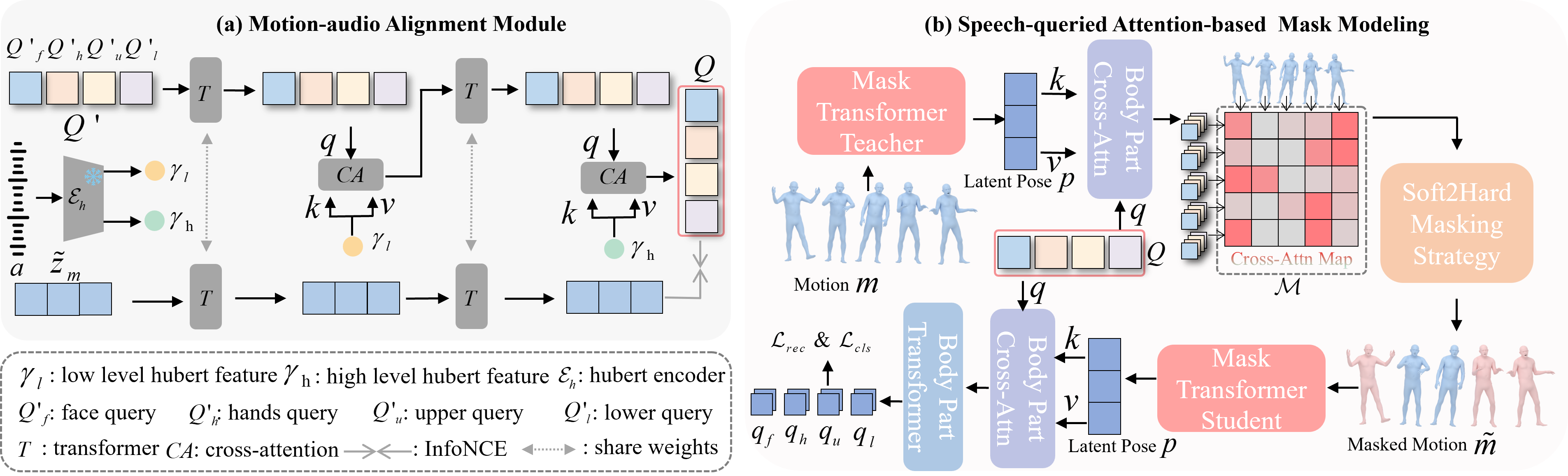}
    \caption{\textbf{Architecture of EchoMask.} \textbf{(a)} MAM projects motion and audio into a shared latent space. Learnable speech queries \(Q'\) are refined through hierarchical cross-attention with HuBERT features (\( \gamma_l, \gamma_h \)) and jointly processed with quantized latent motion \( \tilde{z}_m \) via a shared transformer, optimized with contrastive loss. \textbf{(b)} Given \( m \), mask transformer teacher computes a cross-attention map \( \mathcal{M} \) between latent poses \(p\) and motion-aligned speech features \( Q \), identifying semantically important frames. These frames are masked via a Soft2Hard strategy to produce \( \tilde{m} \), which the student transformer uses to generate motion tokens. }

    \vspace{-15pt}
    \label{fig:method}
\end{figure*}
As illustrated in Figure~\ref{fig:overview}, EchoMask consists of two core modules that work together to enable semantically grounded co-speech motion generation: (1) a hierarchical motion-audio alignment module (MAM) for generating motion-aligned speech features, and (2) a speech-queried attention mechanism (SQA) for identifying and masking key motion frames.

Given input audio \( a \) and quantized latent motion tokens \( z_m \), MAM \( \mathcal{F}_{\text{moau}} \) first projects paired motion and speech features into a shared latent space by leveraging both low-level and high-level HuBERT features. Learnable speech queries interact with these features via cross-attention and self-attention, resulting in motion-aligned speech representations \( Q \):

\begin{equation}
\mathcal{F}_{\text{moau}}: (a, z_m, Q'; \theta_{\text{moau}}) \mapsto Q,
\end{equation}

where \( \theta_{\text{moau}} \) denotes MAM’s parameters and \( Q \) is used to guide the masking process and motion generation.

To identify semantically important frames, a Transformer Teacher \( \mathcal{F}' \) (an EMA copy of the Student \(\mathcal{F}\)) computes a cross-attention map \( \mathcal{M} \) between motion input \( m \) and speech features \( Q \). 

The \( \mathcal{M} \) is then passed to the Soft2Hard masking module \( P(\cdot) \), which masks the frames with high attention scores to yield the masked motion sequence \( \tilde{m} \):

\begin{equation}
\tilde{m} = P(m, \mathcal{M}).
\end{equation}

The fused representation \( q_m \), obtained from \( \mathcal{F}\), \(Q\) and \(\tilde{m}\), is then decoded into the final motion \( \hat{m} \) using an RVQ-VAE decoder \cite{zeghidour2021soundstream,borsos2023audiolm,guo2024momask}. To further disentangle motion semantics across the body, we adopt a part-wise decoder structure following \cite{ao2023gesturediffuclip, liu2024emage}, dividing the body into face, hands, upper body, and lower body segments.

\subsection{Motion-Audio Joint Embedding Learning}  
Previous methods \cite{chen2024diffsheg,liu2024emage,chen2024enabling,yi2023generating} directly take static speech features as conditioning inputs without considering the huge gap between motion and audio modality. This often leads to suboptimal performance due to misalignment between the speech and motion modalities, making it difficult for generative models to effectively bridge the gap. To address this, we propose a hierarchical joint embedding module for motion-audio alignment (MAM) as shown in Figure \ref{fig:method} (a) that projects motion and audio into a shared latent space to generate motion-aligned speech features. The entire module can be divided into two key stages: \textit{Hierarchical Speech Query Encoding} and \textit{Latent Space Alignment}.

\noindent\textbf{Hierarchical Speech Query Encoding.}  
MAM first focuses on refining learnable speech queries through hierarchical feature fusion. We extract both low-level and high-level audio features from a pretrained HuBERT model \cite{hsu2021hubert}, denoted as \( \gamma_l \) and \( \gamma_h \), respectively. These features provide complementary information—phonetic timing from shallow layers and semantic context from deeper layers. Learnable speech queries \( Q'\) are initialized randomly and progressively refined by interacting with these hierarchical speech features through cross-attention layers. The first cross-attention layer fuses local acoustic details, while the second contextualizes them using global semantics. The output \( Q' \) serves as the refined set of speech queries for motion alignment.

\noindent\textbf{Latent Space Alignment.}  
After encoding, the refined speech queries \( Q' \) are aligned with the quantized latent motion sequence \( \tilde{z}_m\) using a shared Transformer \( T(\cdot) \). Both streams are passed through the same transformer layers to ensure feature consistency under identical inductive biases. This shared backbone allows the queries and motion tokens to be co-trained and updated via synchronized gradients. Formally:
\begin{equation}
\hat{Q}, \hat{z}_m = T(Q'), T(\tilde{z}_m),
\end{equation}
where \( \hat{Q} \) and \( \hat{z}_m \) denote the final embeddings of speech queries and motion tokens, respectively.

To enforce tight alignment between the modalities, we introduce a contrastive loss based on InfoNCE at both the frame level and the global (sentence) level. For a batch of \( B \) paired sequences, we define:
\begin{equation}
\mathcal{L}_{\text{align}} = -\sum_{i=1}^{B} \log \frac{\exp\left(\text{sim}(\hat{q}_i, \hat{z}_i)/\tau\right)}{\sum_{j=1}^{B} \exp\left(\text{sim}(\hat{q}_i, \hat{z}_j)/\tau\right)}.
\end{equation}
where \( \hat{q}_i \) and \( \hat{z}_i \) are either the frame-level or pooled sentence-level embeddings for the \( i \)-th pair, \( \text{sim}(\cdot, \cdot) \) is a similarity function (e.g., cosine similarity), and \( \tau \) is a temperature hyperparameter. This contrastive objective ensures that motion and speech embeddings corresponding to the same input are drawn closer together, while embeddings from different pairs are pushed apart. 

\subsection{Speech-queried Attention Mechanism} 
While it may appear reasonable to assume that semantic content is uniformly distributed across a motion sequence, which would make random masking seem effective, this assumption rarely holds in practice. In reality, semantic information tends to be concentrated in specific frames that align closely with speech. Loss-based masking methods \cite{jeonghgm3} assume that frames with higher reconstruction loss are more semantically meaningful. However, in motion generation, these approaches often highlight denser frames rather than truly semantic ones, leading to suboptimal guidance. To better reflect the actual distribution of meaningful content, we introduce a speech-queried attention mechanism (SQA), as shown in Figure \ref{fig:method} (b). The key idea is that semantically important elements naturally draw attention when observing motion. By incorporating attention between motion-aligned speech features and motion sequence, our method selectively identifies and masks the frames most relevant to the spoken content.

\noindent\textbf{Speech-queried Cross-Attention.}
Given the motion-aligned speech features \( Q \) from the MAM and an input motion sequence \(m\), we first obtain its latent pose representation \( p \). The body-part cross-attention module then computes a cross-attention map \( \mathcal{M} \) using \( p \) as keys and \( Q \) as query, processed by a mask transformer teacher. Since the motion is factorized into four regions—face, hands, upper body, and lower body—the speech queries are also projected into a part-aware latent space, enabling alignment between each body part and the speech. The resulting attention map, defined as \( \mathcal{M} = \sum{\mathcal{M}_{\text{parts}}} \), where \( \mathcal{M}_{\text{parts}} \in \{\mathcal{M}_{\text{face}}, \mathcal{M}_{\text{hands}}, \mathcal{M}_{\text{upper}}, \mathcal{M}_{\text{lower}}\} \), captures fine-grained semantic relevance between speech and motion. This facilitates precise identification of which motion frames and parts are most aligned with the speech content. To derive a frame-level importance score \( s \in \mathbb{R}^{T} \), we aggregate attention scores across all motion-aligned speech features \(Q\):
\begin{equation}
s_j = \sum_{i=1}^{T} \mathcal{M}_{i,j}, \quad j \in \{1, \dots, T\},
\end{equation}
where \( s_j \) represents the semantic significance of the \( j \)-th motion frame with respect to the motion-aligned speech features. Frames with higher \( s_j \) values are considered semantically richer and more relevant to speech dynamics.

To promote alignment with semantically meaningful motion, we apply soft supervision \(\mathcal{L}_{sem}\)to the Student’s frame-level attention scores using binary cross-entropy loss against soft or binary semantic labels. This auxiliary objective guides the Student to focus on important frames, and through EMA updates, the Teacher gradually inherits this behavior, yielding more interpretable and speech-consistent attention maps for guiding masking.

\noindent\textbf{Soft-to-Hard Masking Strategy.}  
To avoid prematurely masking high semantic frames, which can hinder effective learning in the early stages, we employ a \textit{soft-to-hard} masking strategy that aligns with the model’s learning progression. The \texttt{argsort}-based masking, which deterministically selects the top semantic frames, is treated as a ``hard'' masking strategy. In contrast, the ``soft'' variant samples frames based on a probability distribution defined by \( s_j \), allowing for stochastic selection.

During training, we gradually shift from soft to hard masking by adjusting their proportions over epochs. Specifically, at training epoch \( t \), the soft and hard mask ratios are updated as:
\begin{equation}
\begin{aligned}
\alpha_t^s &= \alpha_0^s + \frac{t}{T}(\alpha_T^s - \alpha_0^s), \\
\alpha_t^h &= \alpha_0^h - \frac{t}{T}(\alpha_0^h - \alpha_T^h), \\
\alpha_t^r &= \alpha - \alpha_t^s - \alpha_t^h.
\end{aligned}
\end{equation}
where \( \alpha_t^s \), \( \alpha_t^h \), and \(\alpha_t^r\) denote the proportions of soft, hard, and random masks at epoch \( t \), and \( \alpha_0^s, \alpha_T^s, \alpha_0^h, \alpha_T^h \) are predefined initial and final values. \(\alpha\) is mask ratio. 

\subsection{Inference}
\begin{figure}
    \centering
    \includegraphics[width=0.49\textwidth]{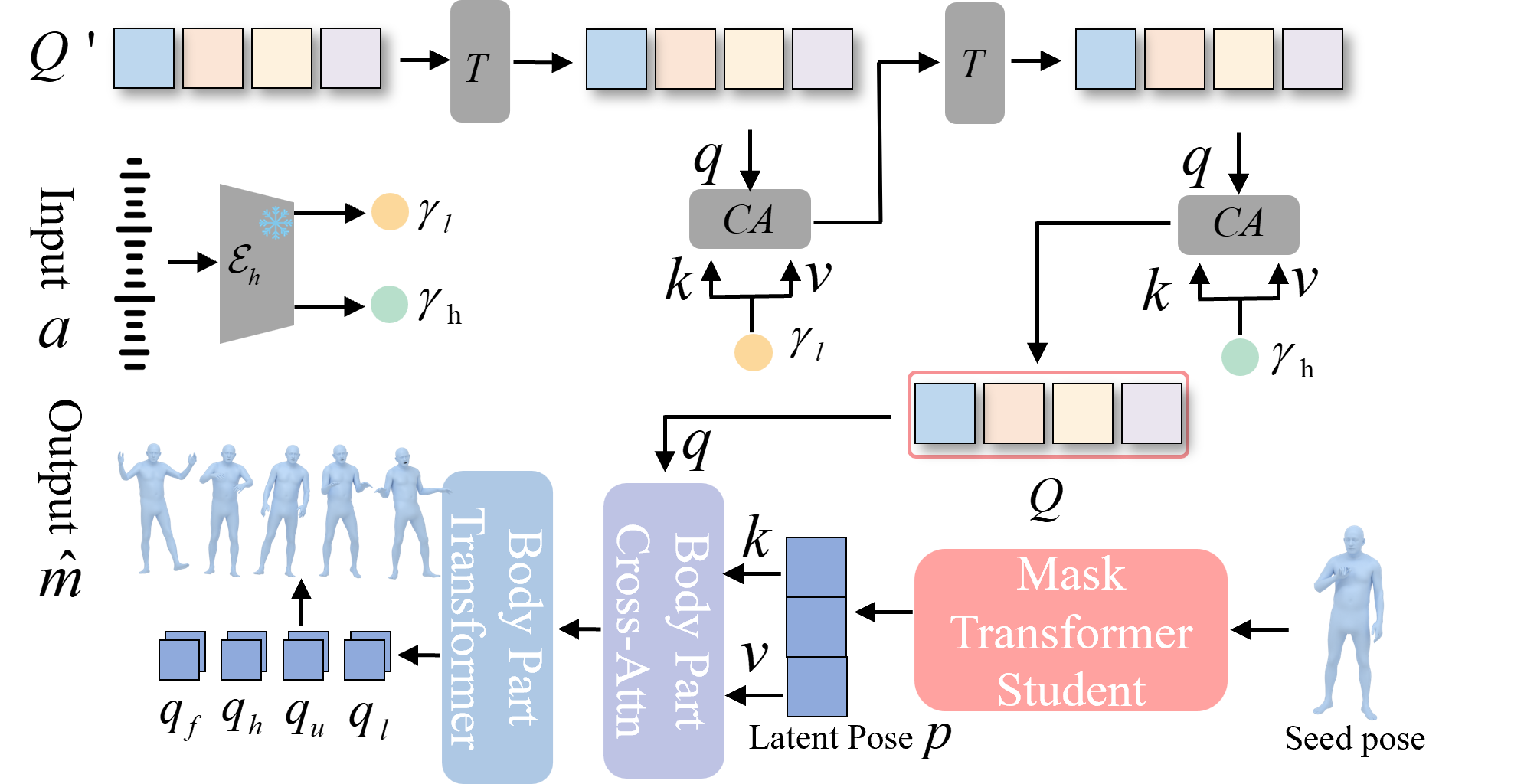}
    \caption{Pipeline of Inference. EchoMask takes speech as the sole input and employs hierarchical HuBERT features to guide a mask transformer, initialized with four seed pose frames. The model predicts motion tokens with the guidance of motion-aligned speech features \(Q\) that are subsequently decoded by the RVQ-VAE decoder to generate whole-body motion.
}
    \vspace{-15pt}
    \label{fig:inference}
\end{figure}

As shown in Figure~\ref{fig:inference}, EchoMask generates co-speech motion using speech input alone. Unlike prior methods that directly use features from a pretrained audio encoder as static conditions, we leverage motion-aligned speech representations to guide generation. Specifically, hierarchical speech features are first extracted from a pretrained HuBERT encoder \( \mathcal{E}_h \), then fused with learnable speech queries via cross-attention in the MAM. This results in refined, motion-aligned speech features \( Q \), which capture both low-level prosody and high-level semantic intent aligned with the motion.

Using only four seed motion frames and the aligned speech features \( Q \), we initialize the masked motion sequence and input it to the transformer student. The model predicts the full sequence of motion tokens, which are then decoded by the RVQ-VAE decoder to generate the final co-speech motion.

\section{Experiments}
\subsection{Experimental Setup}
\noindent\textbf{Datasets.} 
For training and evaluation, we use the BEAT2 dataset \cite{liu2024emage}, which contains 60 hours of high-quality finger motion data from 25 speakers (12 female, 13 male). The dataset includes 1762 sequences, each averaging 65.66 seconds, where speakers respond to daily inquiries. We divide the dataset into training (85\%), validation (7.5\%), and test (7.5\%) sets. To ensure a fair comparison, we follow \cite{liu2024emage} and use data from Speaker 2 for training and validation.

\noindent \textbf{Implementation Details.} 
Our model is trained on a single NVIDIA A100 GPU for 200 epochs with a batch size of 64. The  RVQ-VAE is downscaled by 4. The residual quantization has 6 layers, a codebook size of 256, and a dropout rate of 0.2. The training uses the ADAM optimizer with a 1e-4 learning rate. During inference, we use a four-frame seed pose to initialize each motion clip. For consecutive clips, the last four frames of the previous segment are overlapped and reused as the seed for the next, following \cite{liu2024emage}.

\begin{figure*}
    \centering
    \includegraphics[width=0.9\textwidth]{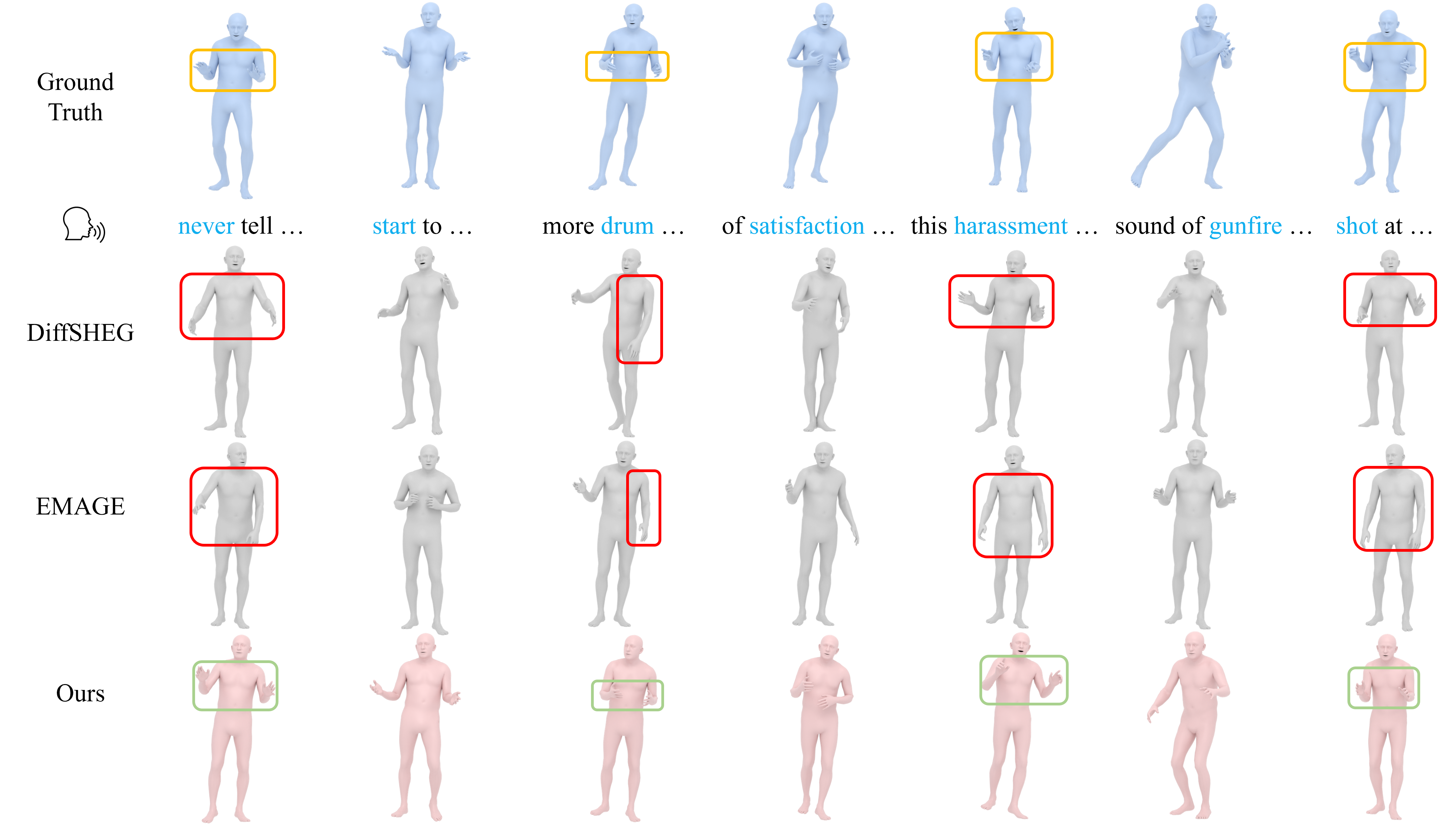}
    \caption{Visual comparison.  Red boxes highlight implausible or uncoordinated motions, while green boxes indicate coherent and semantically appropriate results. Our EchoMask consistently generates co-speech motions that are semantically aligned with ground truth. For instance, when articulating “never” and “start”, our model positions both hands in a poised gesture near the torso, reflecting a thoughtful and intentional motion, whereas prior methods such as DiffSHEG and EMAGE either generate imbalanced hand postures or ambiguous limb placements.}
    \vspace{-15pt}
    \label{fig:compare}
\end{figure*}
\begin{figure}
    \centering
    \includegraphics[width=0.49\textwidth]{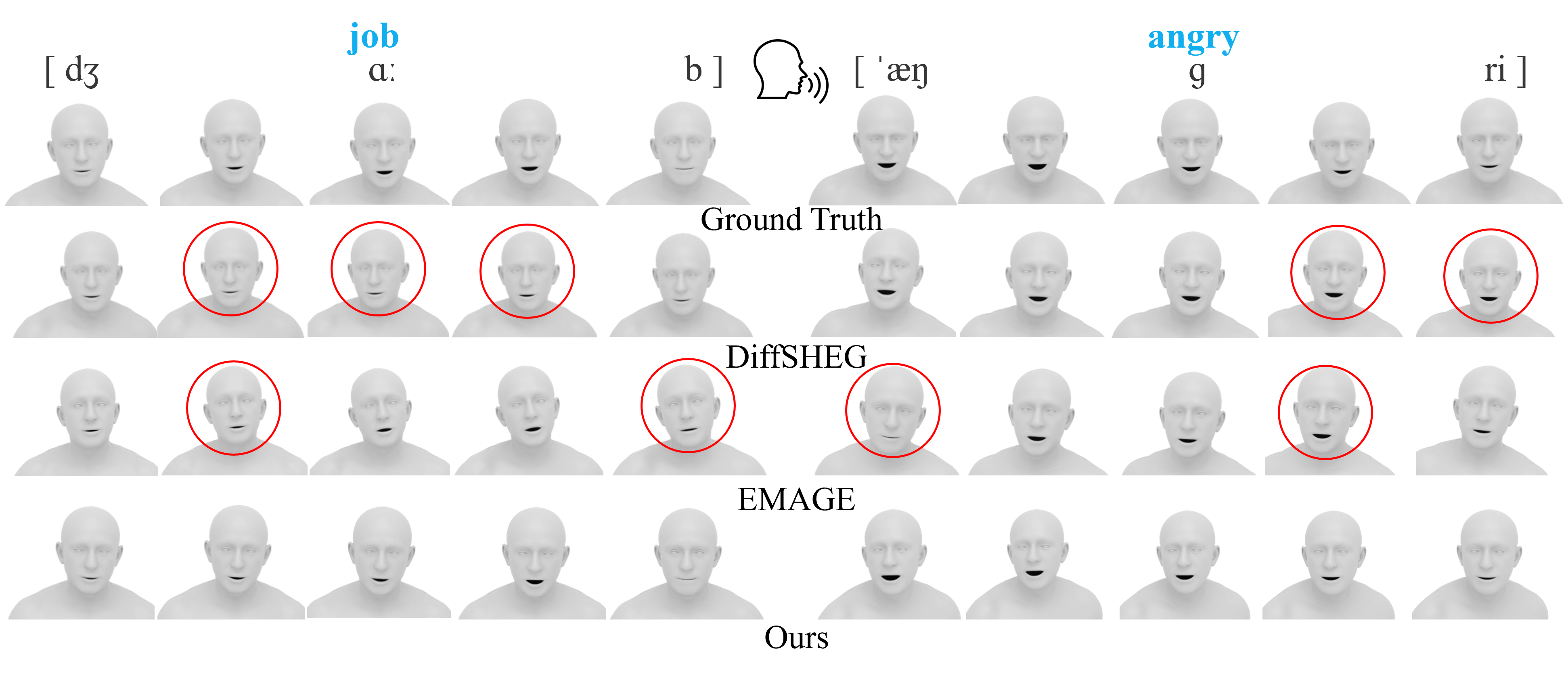}
    \caption{Facial Comparison.}
    \vspace{-15pt}
    \label{fig:face}
\end{figure}

\noindent \textbf{Metrics.}
We assess the quality of generated body gestures using the FGD metric \cite{yoon2020speech}, which evaluates how closely the distribution of generated gestures aligns with ground truth (GT), providing a measure of realism. Gesture diversity is quantified using DIV \cite{li2021audio2gestures}, calculated as the average L1 distance across multiple gesture clips to capture motion variation. To evaluate speech-motion synchrony, we employ BC \cite{li2021ai}, which measures the temporal alignment between gesture rhythm and audio beats. For facial expression accuracy, we use two reconstruction metrics: vertex MSE \cite{yang2023diffusestylegesture} to assess positional differences and vertex L1 difference (LVD) \cite{yi2023generating} to quantify discrepancies between GT and generated facial vertices. Additionally, we conduct a user study to provide a more comprehensive evaluation.
\subsection{Qualitative Results}
\noindent\textbf{Qualitative Comparisons.} As illustrated in Figure~\ref{fig:compare}, our EchoMask consistently produces co-speech motions that are both semantically aligned with the ground truth and physically expressive. In contrast, baseline methods such as DiffSHEG and EMAGE often generate gestures that are misaligned with the underlying speech semantics and visually implausible.

In the case of “drum”, EchoMask captures the rhythmic semantics by swinging one arm outward in a dynamic arc, a detail that other methods miss—often producing static or downward-pointing arms. Similarly, for the term “satisfaction”, our method aligns the gesture with the meaning by raising the right arm close to the chest in a self-reflective manner, while other baselines lack such subtlety.
\sloppy  
Interestingly, when expressing “harassment” and “gunfire”, EchoMask emphasizes the tension through body posture—one arm bent with visible muscular contraction and the upper body leaning slightly forward—accurately conveying urgency or defensive response. Competing DiffSHEG and EMAGE display either stiffness or lack of spatial coordination in these challenging cases.

For the word “shot”, our model captures the implied action with a firm forward-facing hand posture, where both arms hold an assertive position, demonstrating a level of contextual understanding missing from DiffSHEG and EMAGE, which often place hands too low or render them passively. Throughout all these cases, EchoMask not only maintains consistency in hand dominance and articulation range but also delivers a broad repertoire of movements that better align with the acoustic-semantic cues of the spoken words.

This qualitative evidence underscores the strength of our speech-queried attention mask modeling in driving expressive and semantically grounded motion synthesis.

For facial comparison, as shown in Figure~\ref{fig:face}, red circles highlight mismatches or unnatural expressions from baseline methods, which often produce stiff or poorly timed facial motions. In contrast, our method generates smooth and expressive transitions that better align with the phonetic structure and emotional tone of speech. Key articulatory dynamics—like lip closure for \textipa{[{\ae}N]} and jaw opening for \textipa{[dZ]}—are more accurately captured, demonstrating the effectiveness of our part-aware, speech-aligned facial synthesis.

\begin{figure}
    \centering
    \includegraphics[width=0.43\textwidth]{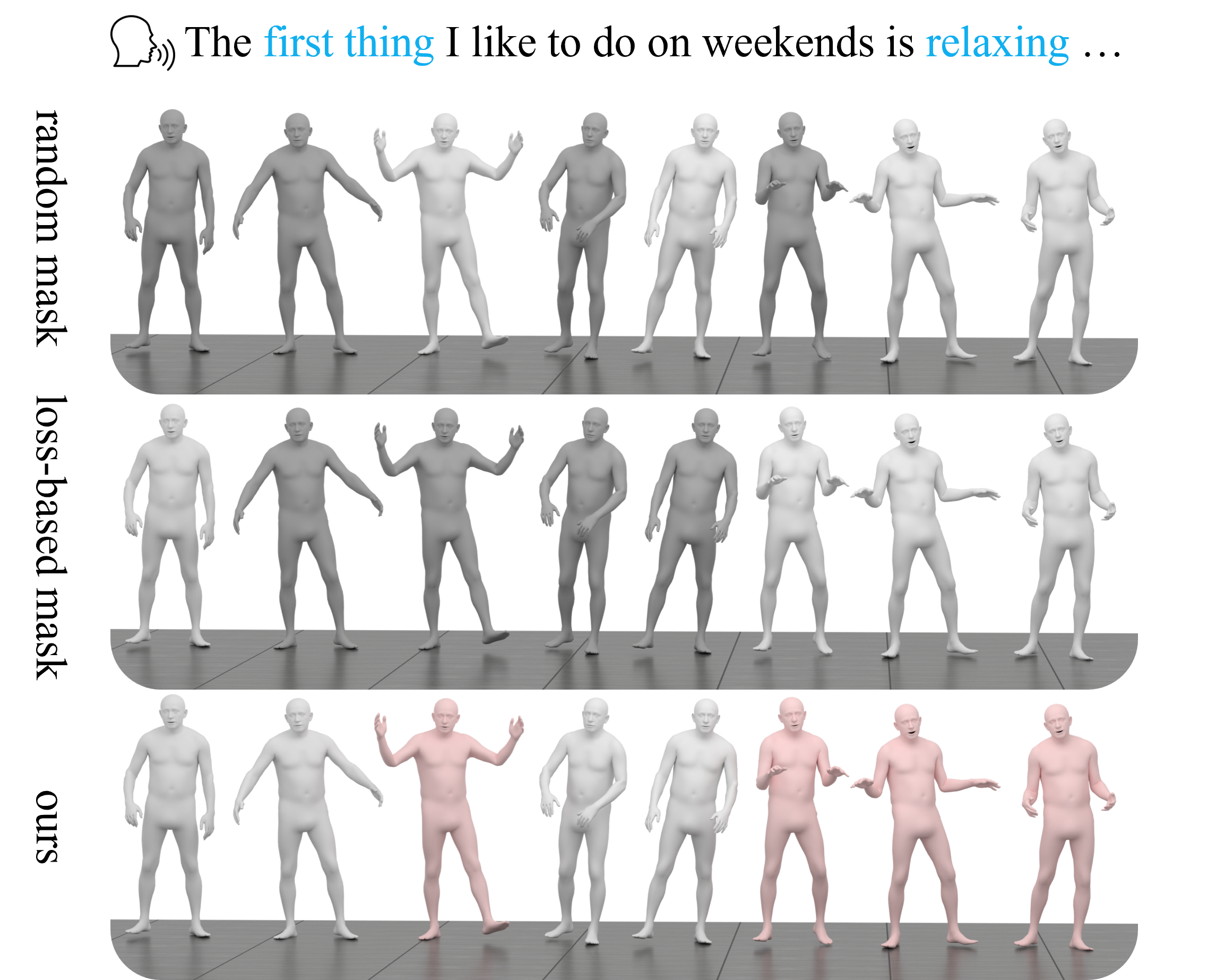}
    \caption{Visualization of generated masked motion by random mask, loss-based mask, and our method. The darker and red motion frames represent those that are masked out.}
    \vspace{-15pt}
    \label{fig:strategy}
\end{figure}

\noindent\textbf{Masking Strategy.} As illustrated in Figure~\ref{fig:strategy}, the random masking strategy (top row) produces a uniform, indiscriminate pattern, often masking low-information frames or temporally scattered segments. The loss-based masking strategy (middle row), which selects frames with high reconstruction error, similarly exhibits limitations. While these frames often reflect abrupt pose transitions or motion discontinuities, they do not reliably correspond to semantically salient content—frequently capturing transitional noise or speech pauses rather than meaningful gestures.

In contrast, our speech-queried masking strategy (bottom row) yields a targeted and speech-synchronized masking pattern. The masked frames are concentrated around semantically rich regions, closely aligned with gesture peaks that convey the intent of the spoken words—for example, the emphatic upward motion for “first” and the fluid, relaxed hand movement for “relaxing.” By explicitly grounding the masking in cross-modal attention, our approach ensures the model learns to reconstruct motion segments that are both contextually and semantically significant. 

\begin{figure}
    \centering
    \includegraphics[width=0.45\textwidth]{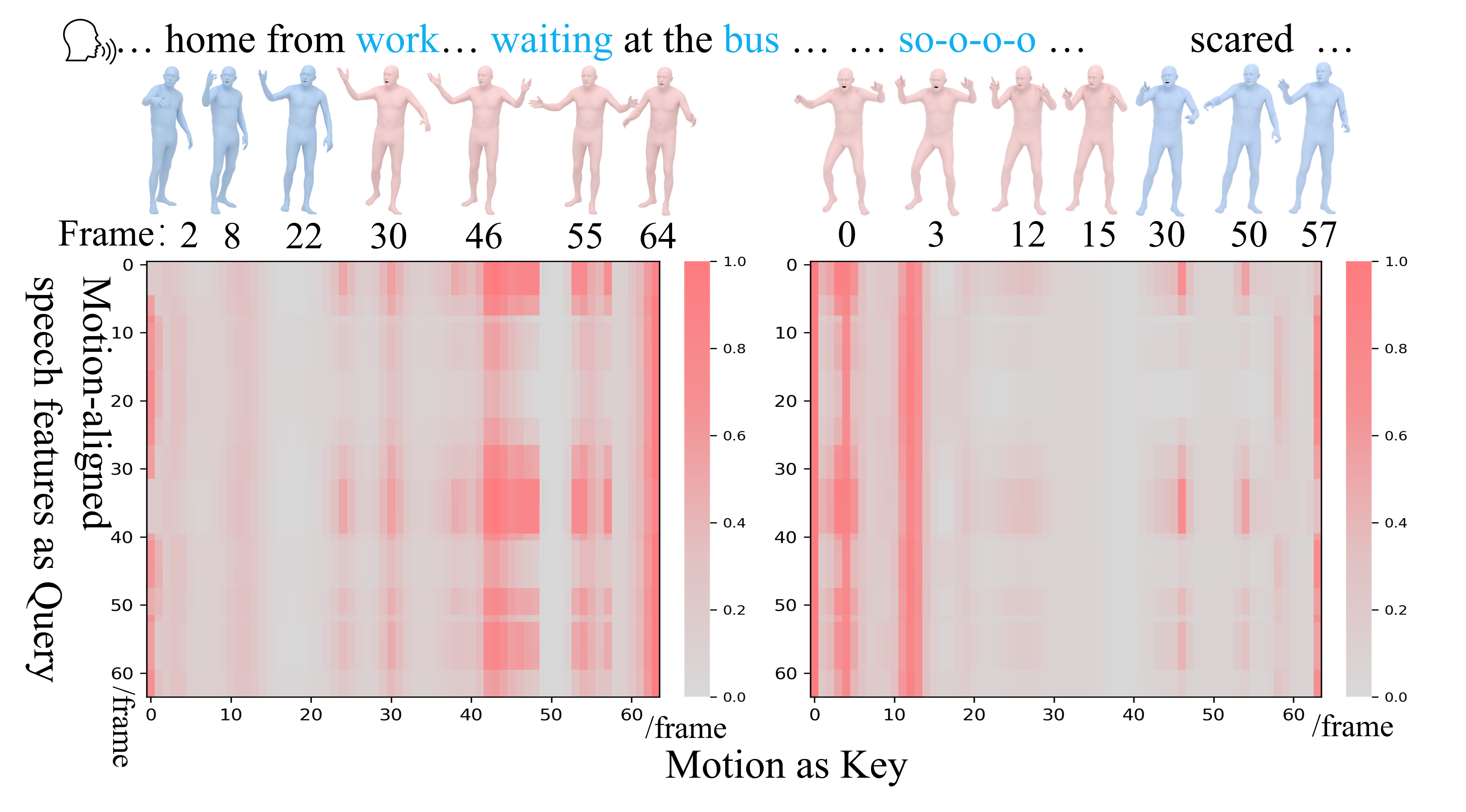}
    \caption{Visualization of the cross-attention map between speech queries and motion frames. Darker red regions indicate higher attention scores. Our model attends to semantically rich motion frames that align closely with key speech tokens such as “work,” “waiting,” and “so-o-o-o,” demonstrating effective cross-modal alignment.
}    \vspace{-15pt}
    \label{fig:attn_map}
\end{figure}

\noindent\textbf{Cross-Attn Map.}
As shown in Figure~\ref{fig:attn_map},  our model assigns high attention scores to semantically expressive regions, such as “work,” “waiting,” and the elongated syllables in “so-o-o-o,” demonstrating its ability to capture both word-level semantics and prosodic emphasis. Unlike previous approaches with diffuse or noisy attention, our speech-queried mechanism produces focused and interpretable patterns. The attention selectively highlights gesture-relevant frames while de-emphasizing idle motions, enabling more effective masking and improving alignment between speech and motion.

\begin{figure}
    \centering
    \includegraphics[width=0.48\textwidth]{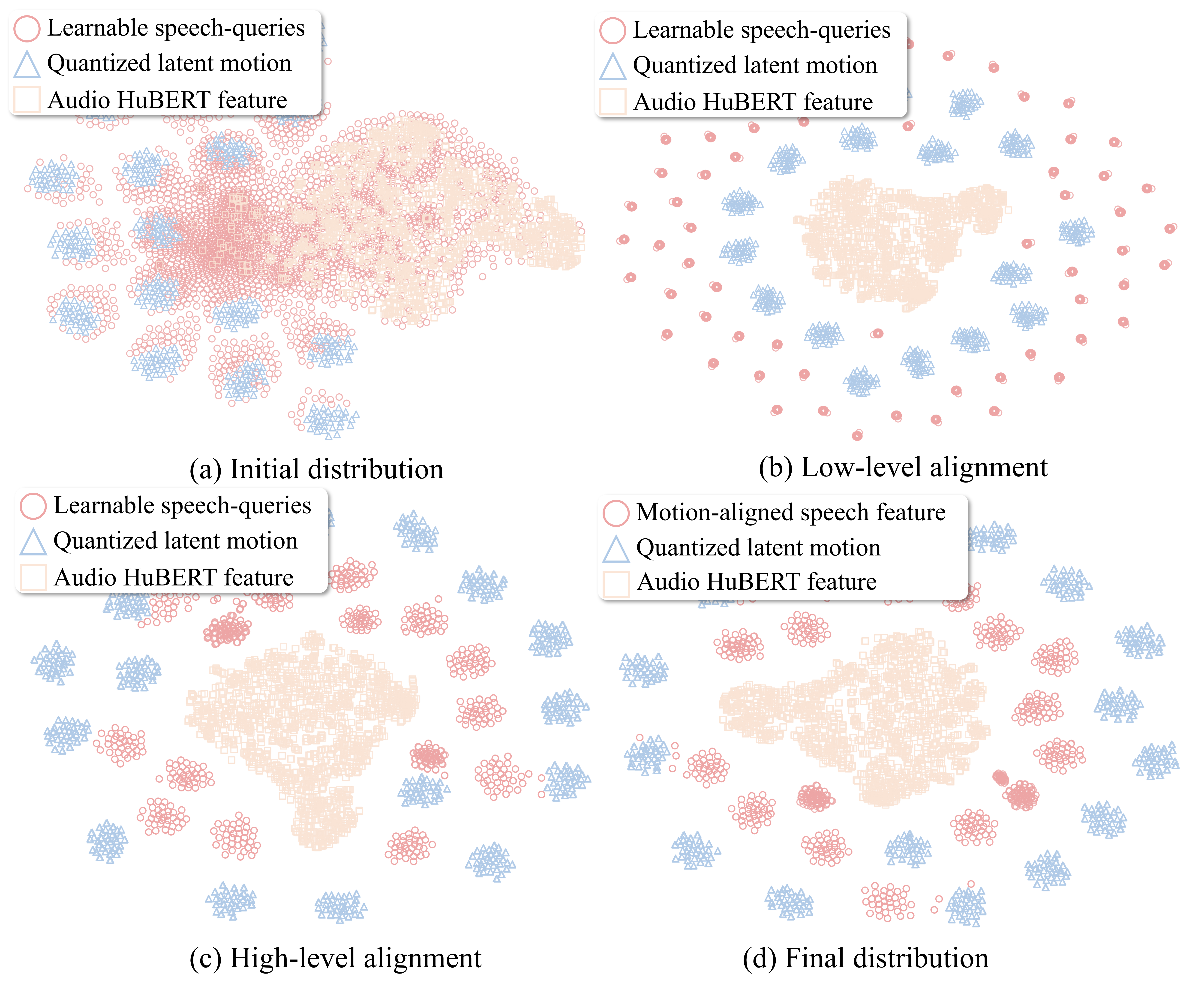}
    \caption{t-SNE \cite{van2008visualizing} visualization of the shared latent space. After MAM, motion-aligned speech features (red circles), quantized latent motion tokens (blue triangles), and HuBERT audio features (peach squares) form well-aligned and semantically coherent clusters, illustrating effective modality fusion and cross-modal alignment.
}
    \vspace{-15pt}
    \label{fig:ablation_moau}
\end{figure}
\begin{figure}
    \centering
    \includegraphics[width=0.4\textwidth]{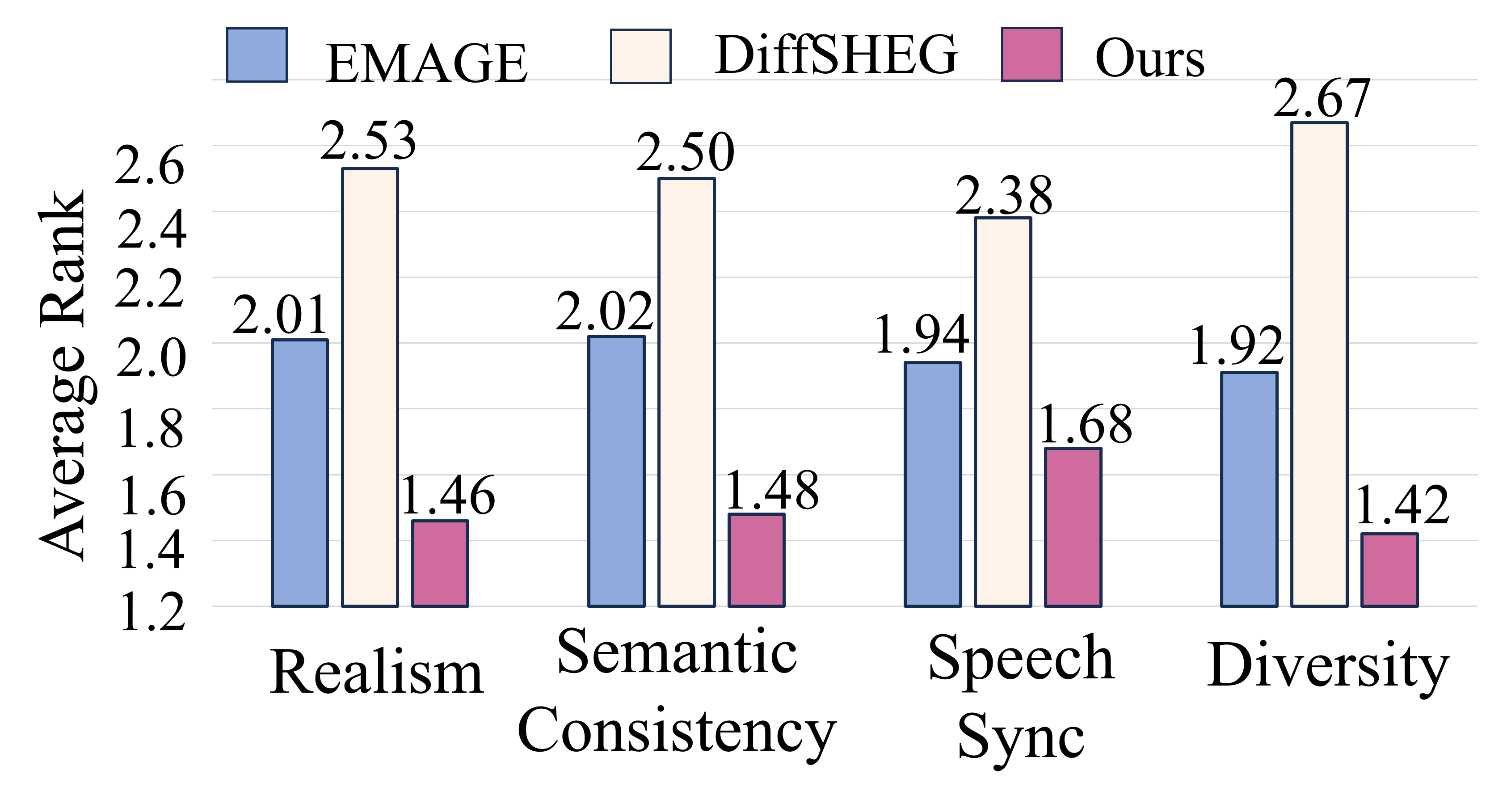}
    \caption{Results of the user study.}
    \vspace{-15pt}
    \label{fig:user}
\end{figure}

\noindent\textbf{The Effectiveness of MAM.}
Figure~\ref{fig:ablation_moau} illustrates the progressive refinement of speech features through our MAM module. Each point represents a frame-level feature. Initially (top-left), the learnable speech queries (red circles) are scattered and poorly aligned with motion tokens (blue triangles) and audio features (peach squares), indicating weak semantic grounding.  After incorporating low-level hierarchical (top-right) and low-level \(\gamma_h\)(bottom-left) HuBERT features \(\gamma_l\), the distribution becomes more structured. In the final stage (bottom-right), the motion-aligned speech features form compact clusters that are well integrated between audio and motion, demonstrating successful alignment in the shared latent space. This alignment improves attention computation and supports semantically coherent and expressive motion generation, validating the effectiveness of each MAM component.

\noindent\textbf{User Study.} To assess perceptual quality, we conducted a user study involving 28 participants with varied backgrounds, who evaluated 10 videos across multiple methods. The evaluation focused on four aspects: realism, alignment with speech semantics, temporal synchrony between motion and speech, and diversity of motion. As illustrated in \ref{fig:user}, our method was consistently rated highest.

\subsection{Quantitative Results}
\begin{table}[t] 
   \centering
    \resizebox{0.45\textwidth}{!}{\begin{tabular}{lccccc}
        \toprule
         \textbf{Method} & \textbf{FGD$\downarrow$} & \textbf{BC$\uparrow$} & \textbf{DIV$\uparrow$} & \textbf{MSE$\downarrow$} & \textbf{LVD$\downarrow$} \\
        \midrule
        \rowcolor{gray!20}
        \multicolumn{6}{l}{\textbf{Facial Generation}} \\
        \cmidrule(lr){1-6}
         FaceFormer~\cite{fan2022faceformer} & - & - & - & 7.787 & 7.593 \\
         CodeTalker~\cite{xing2023codetalker} & - & - & - & 8.026 & 7.766 \\
         \cmidrule(lr){1-6}
         \rowcolor{gray!20}
        \multicolumn{6}{l}{\textbf{Non-facial Gesture Generation}} \\
        \cmidrule(lr){1-6}
        DisCo~\cite{liu2022disco} & 9.680 & 6.441 & 9.892 & - & - \\
        HA2G~\cite{liu2022learning} & 12.14 & 6.711 & 8.916 & - & - \\
         CaMN~\cite{liu2022beat} & 6.644 & 6.769 & 10.86 & - & - \\
        LivelySpeaker~\cite{habibie2021learning} & 11.80 & 6.659 & 11.28 & - & - \\
         DSG~\cite{yang2023diffusestylegesture} & 8.811 & 7.241 & 11.49 & - & - \\
         \cmidrule(lr){1-6}
         \rowcolor{gray!20}
         \multicolumn{6}{l}{\textbf{Holistic Motion Generation}} \\
        \cmidrule(lr){1-6}
         Habibie \textit{et al.}~\cite{habibie2021learning} & 9.040 & 7.716 & 8.213 & 8.614 & 8.043 \\
         TalkSHOW~\cite{yi2023generating} & 6.209 & 6.947 & \textbf{13.47} & 7.791 & 7.771 
        \\
         EMAGE~\cite{liu2024emage} & 5.512 & 7.724 & 13.06 & 7.680 & 7.556 \\
         DiffSHEG~\cite{chen2024diffsheg} & 8.986 & 7.142 & 11.91 & 7.665 & 8.673 
        \\
        \cmidrule(lr){1-6}
         \textbf{EchoMask (Ours)} & \textbf{4.623} & \textbf{7.738} & 13.37 & \textbf{6.761} & \textbf{7.290} \\
        \bottomrule
    \end{tabular}}
    \caption{\textbf{Quantitative comparison with SOTA.} Lower values indicate better performance for FMD, FGD, MSE, and LVD, while higher values are better for BC and DIV. For clarity, we report FGD \(\times 10^{-1}\), BC \(\times 10^{-1}\), MSE \(\times 10^{-8}\), and LVD \(\times 10^{-5}\). Best results are shown in \textbf{bold}.}
    \vspace{-10pt}
    
    \label{tab:compare1}
\end{table}
\noindent\textbf{Comparison with Baselines.}
Table~\ref{tab:compare1} presents a comprehensive quantitative comparison between EchoMask and a wide range of state-of-the-art methods across facial, non-facial, and holistic motion generation tasks. Our method consistently achieves the best performance across almost all metrics. In holistic motion generation, EchoMask outperforms all baselines with the lowest FGD, MSE, and LVD, indicating superior realism, motion accuracy, and temporal smoothness. It also achieves the highest BC, reflecting better rhythm alignment with speech. While TalkSHOW attains the highest DIV, EchoMask remains highly competitive, suggesting it captures a broad range of expressive motions without sacrificing structure or semantic fidelity. In facial generation, EchoMask surpasses strong baselines such as FaceFormer and CodeTalker with significant reductions in MSE and LVD, demonstrating more accurate and emotionally coherent facial articulation.

\noindent\textbf{Ablation Study on Components.}
As Table~\ref{tab:compare1} shows, replacing the VQ-VAE with the residual variant (RVQ-VAE) results in consistent improvements across metrics. Incorporating our speech-queried attention mask modeling (SQA) leads to further gains. While random and loss-based masking offer moderate improvements, the attention-based approach achieves the lowest FGD and highest motion diversity, highlighting its ability to identify semantically informative frames aligned with the speech content. We also examine the impact of the MAM design. Using either low-level or high-level HuBERT features in isolation yields competitive results; however, fusing both levels significantly enhances motion diversity and beat consistency. Moreover, excluding the alignment loss \(\mathcal{L}_{\text{align}}\) causes a clear degradation in both FGD and DIV, underscoring its role in maintaining cross-modal consistency. The complete MAM configuration delivers the most balanced performance across all metrics. Finally, the full EchoMask model outperforms all ablated variants, validating the effectiveness of the proposed SQA and MAM.

\begin{table}[t] 
   \centering
   \small
    \resizebox{0.48\textwidth}{!}{\begin{tabular}{lccccc}
        \toprule
         \textbf{Method} & \textbf{FGD$\downarrow$} & \textbf{BC$\uparrow$} & \textbf{DIV$\uparrow$} & \textbf{MSE$\downarrow$} & \textbf{LVD$\downarrow$} \\
        \midrule

         EchoMask(VQ-VAE) & 6.664 & 7.464 & 10.86 & 7.225 & 7.693 \\
         + RVQ-VAE & 6.106 & 7.654 & 11.68 & 7.014 & 7.484 \\
         \cmidrule(lr){1-6}
         \rowcolor{gray!20}
        \multicolumn{6}{l}{\textbf{+ Speech-queried Attention Mechanism}} \\
        \cmidrule(lr){1-6}
        SQA (random mask) & 5.889 & 7.613 & 12.26 & 7.122 & 7.473 \\
        SQA (loss-based mask) & 5.745 & 7.607 & 12.66 & 7.056 & 7.454\\
        SQA (ours) & 5.455 & 7.615 & 12.83 & 6.976 & 7.364 \\
         \cmidrule(lr){1-6}
         \rowcolor{gray!20}
         \multicolumn{6}{l}{\textbf{+ Hierarchical Motion-audio Alignment}} \\
        \cmidrule(lr){1-6}
         MAM (low-level feature) & 5.679 & 7.647 & 13.15 & 7.153 & 7.478 \\
         MAM (high-level feature) & 5.442 & 7.692 & 13.02 & 6.903 & 7.359 \\
        MAM (low\&high-level feature) & 5.420 & 7.684 & 13.21 & 6.977 & 7.346 
        \\
        MAM (\(\mathcal{L}_{align}\)) & 5.887 & 7.571 & 12.78 & 6.988 & 7.393 \\
        MAM (ours) & 5.029 & 7.690 & 13.32 & 6.827 & 7.310
        \\
        \cmidrule(lr){1-6}
         \textbf{EchoMask (Ours)} & \textbf{4.623} & \textbf{7.738} & \textbf{13.37} & \textbf{6.761} & \textbf{7.290} \\
        \bottomrule
    \end{tabular}}
    \caption{Ablation study evaluating the effectiveness of each component within the EchoMask.}
    \vspace{-10pt}
    
    \label{tab:compare1}
\end{table}
\begin{figure}
    \centering
    \includegraphics[width=0.43\textwidth]{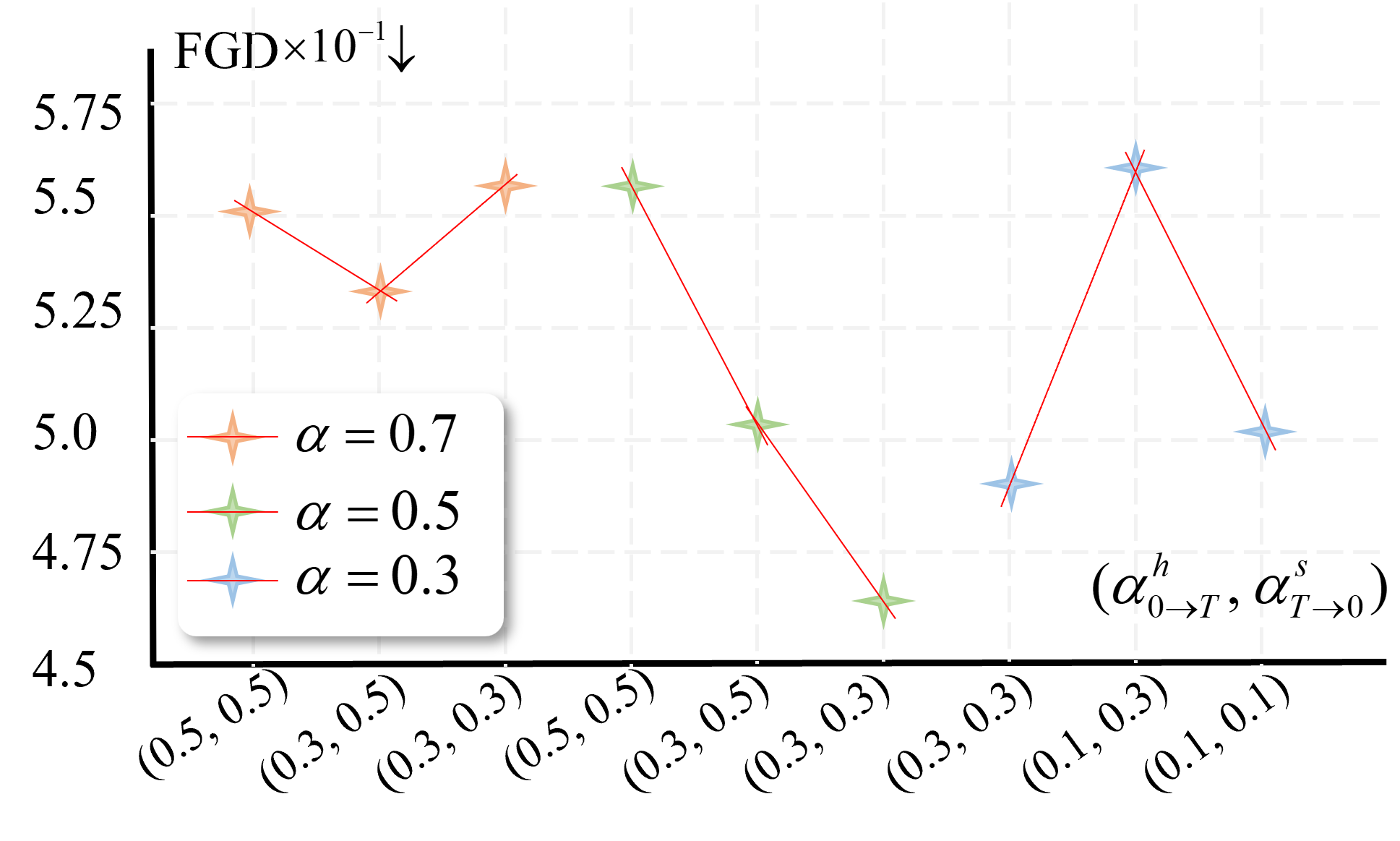}
    \caption{Ablation study on the soft-to-hard masking strategy. We analyze the impact of varying the masking ratios: $\alpha_0^{h}$, $\alpha_{0\to T}^{h}$ for hard masks, and $\alpha_0^{s}$, $\alpha_{T\to 0}^{s}$ for soft masks.}
    \vspace{-15pt}
    \label{fig:s2h}
\end{figure}

\noindent\textbf{Soft-to-hard Masking Strategy.}  
Figure ~\ref{fig:s2h} reports an ablation study examining the impact of various soft-to-hard masking schedules on generation quality, evaluated by FGD. The results indicate that a balanced configuration (\(\alpha = 0.5\)) with a smooth transition from soft to hard masking (\(\alpha_0^h = 0, \alpha_T^h = 0.3\); \(\alpha_0^s = 0.3, \alpha_T^s = 0\)) yields the best performance. This highlights the advantage of starting with probabilistic frame sampling, which encourages broader motion exploration, and progressively shifting to deterministic selection based on attention, allowing the model to focus on semantically salient frames as training advances.

\section{Conclusion}
In this work, we propose EchoMask, a new masked motion modeling framework for holistic co-speech motion generation. EchoMask identifies semantically expressive co-speech motions using motion-aligned speech features, facilitating effective mask modeling in training. We introduce two key components in EchoMask: a motion-audio alignment module and a speech-queried attention mechanism. The former is a hierarchical cross-modal alignment module that embeds motion and audio into a unified latent space through shared attention and contrastive learning. The latter utilizes motion-aligned speech queries to dynamically identify and mask semantically rich motion frames, improving the model’s ability to learn meaningful speech-conditioned motion patterns. Through extensive experiments, EchoMask demonstrates state-of-the-art performance in facial, gestural, and whole-body motion generation. The qualitative results further show its capacity to generate temporally coherent, diverse, and speech-synchronized motions.

\bibliographystyle{ACM-Reference-Format}

\begin{thebibliography}{59}


\ifx \showCODEN    \undefined \def \showCODEN     #1{\unskip}     \fi
\ifx \showISBNx    \undefined \def \showISBNx     #1{\unskip}     \fi
\ifx \showISBNxiii \undefined \def \showISBNxiii  #1{\unskip}     \fi
\ifx \showISSN     \undefined \def \showISSN      #1{\unskip}     \fi
\ifx \showLCCN     \undefined \def \showLCCN      #1{\unskip}     \fi
\ifx \shownote     \undefined \def \shownote      #1{#1}          \fi
\ifx \showarticletitle \undefined \def \showarticletitle #1{#1}   \fi
\ifx \showURL      \undefined \def \showURL       {\relax}        \fi
\providecommand\bibfield[2]{#2}
\providecommand\bibinfo[2]{#2}
\providecommand\natexlab[1]{#1}
\providecommand\showeprint[2][]{arXiv:#2}

\bibitem[Ahuja et~al\mbox{.}(2022)]%
        {ahuja2022low}
\bibfield{author}{\bibinfo{person}{Chaitanya Ahuja}, \bibinfo{person}{Dong~Won Lee}, {and} \bibinfo{person}{Louis-Philippe Morency}.} \bibinfo{year}{2022}\natexlab{}.
\newblock \showarticletitle{Low-resource adaptation for personalized co-speech gesture generation}. In \bibinfo{booktitle}{\emph{Proceedings of the IEEE/CVF Conference on Computer Vision and Pattern Recognition}}. \bibinfo{pages}{20566--20576}.
\newblock


\bibitem[Alexanderson et~al\mbox{.}(2023)]%
        {alexanderson2023listen}
\bibfield{author}{\bibinfo{person}{Simon Alexanderson}, \bibinfo{person}{Rajmund Nagy}, \bibinfo{person}{Jonas Beskow}, {and} \bibinfo{person}{Gustav~Eje Henter}.} \bibinfo{year}{2023}\natexlab{}.
\newblock \showarticletitle{Listen, denoise, action! audio-driven motion synthesis with diffusion models}.
\newblock \bibinfo{journal}{\emph{ACM Transactions on Graphics (TOG)}} \bibinfo{volume}{42}, \bibinfo{number}{4} (\bibinfo{year}{2023}), \bibinfo{pages}{1--20}.
\newblock


\bibitem[Ao et~al\mbox{.}(2022)]%
        {ao2022rhythmic}
\bibfield{author}{\bibinfo{person}{Tenglong Ao}, \bibinfo{person}{Qingzhe Gao}, \bibinfo{person}{Yuke Lou}, \bibinfo{person}{Baoquan Chen}, {and} \bibinfo{person}{Libin Liu}.} \bibinfo{year}{2022}\natexlab{}.
\newblock \showarticletitle{Rhythmic gesticulator: Rhythm-aware co-speech gesture synthesis with hierarchical neural embeddings}.
\newblock \bibinfo{journal}{\emph{ACM Transactions on Graphics (TOG)}} \bibinfo{volume}{41}, \bibinfo{number}{6} (\bibinfo{year}{2022}), \bibinfo{pages}{1--19}.
\newblock


\bibitem[Ao et~al\mbox{.}(2023)]%
        {ao2023gesturediffuclip}
\bibfield{author}{\bibinfo{person}{Tenglong Ao}, \bibinfo{person}{Zeyi Zhang}, {and} \bibinfo{person}{Libin Liu}.} \bibinfo{year}{2023}\natexlab{}.
\newblock \showarticletitle{Gesturediffuclip: Gesture diffusion model with clip latents}.
\newblock \bibinfo{journal}{\emph{ACM Transactions on Graphics (TOG)}} \bibinfo{volume}{42}, \bibinfo{number}{4} (\bibinfo{year}{2023}), \bibinfo{pages}{1--18}.
\newblock


\bibitem[Bandara et~al\mbox{.}(2023)]%
        {bandara2023adamae}
\bibfield{author}{\bibinfo{person}{Wele Gedara~Chaminda Bandara}, \bibinfo{person}{Naman Patel}, \bibinfo{person}{Ali Gholami}, \bibinfo{person}{Mehdi Nikkhah}, \bibinfo{person}{Motilal Agrawal}, {and} \bibinfo{person}{Vishal~M Patel}.} \bibinfo{year}{2023}\natexlab{}.
\newblock \showarticletitle{Adamae: Adaptive masking for efficient spatiotemporal learning with masked autoencoders}. In \bibinfo{booktitle}{\emph{Proceedings of the IEEE/CVF Conference on Computer Vision and Pattern Recognition}}. \bibinfo{pages}{14507--14517}.
\newblock


\bibitem[Bao et~al\mbox{.}(2021)]%
        {bao2021beit}
\bibfield{author}{\bibinfo{person}{Hangbo Bao}, \bibinfo{person}{Li Dong}, \bibinfo{person}{Songhao Piao}, {and} \bibinfo{person}{Furu Wei}.} \bibinfo{year}{2021}\natexlab{}.
\newblock \showarticletitle{Beit: Bert pre-training of image transformers}.
\newblock \bibinfo{journal}{\emph{arXiv preprint arXiv:2106.08254}} (\bibinfo{year}{2021}).
\newblock


\bibitem[Borsos et~al\mbox{.}(2023)]%
        {borsos2023audiolm}
\bibfield{author}{\bibinfo{person}{Zal{\'a}n Borsos}, \bibinfo{person}{Rapha{\"e}l Marinier}, \bibinfo{person}{Damien Vincent}, \bibinfo{person}{Eugene Kharitonov}, \bibinfo{person}{Olivier Pietquin}, \bibinfo{person}{Matt Sharifi}, \bibinfo{person}{Dominik Roblek}, \bibinfo{person}{Olivier Teboul}, \bibinfo{person}{David Grangier}, \bibinfo{person}{Marco Tagliasacchi}, {et~al\mbox{.}}} \bibinfo{year}{2023}\natexlab{}.
\newblock \showarticletitle{Audiolm: a language modeling approach to audio generation}.
\newblock \bibinfo{journal}{\emph{IEEE/ACM transactions on audio, speech, and language processing}}  \bibinfo{volume}{31} (\bibinfo{year}{2023}), \bibinfo{pages}{2523--2533}.
\newblock


\bibitem[Cassell et~al\mbox{.}(2001)]%
        {cassell2001beat}
\bibfield{author}{\bibinfo{person}{Justine Cassell}, \bibinfo{person}{Hannes~H{\"o}gni Vilhj{\'a}lmsson}, {and} \bibinfo{person}{Timothy Bickmore}.} \bibinfo{year}{2001}\natexlab{}.
\newblock \showarticletitle{Beat: the behavior expression animation toolkit}. In \bibinfo{booktitle}{\emph{Proceedings of the 28th annual conference on Computer graphics and interactive techniques}}. \bibinfo{pages}{477--486}.
\newblock


\bibitem[Chang et~al\mbox{.}(2023)]%
        {chang2023muse}
\bibfield{author}{\bibinfo{person}{Huiwen Chang}, \bibinfo{person}{Han Zhang}, \bibinfo{person}{Jarred Barber}, \bibinfo{person}{AJ Maschinot}, \bibinfo{person}{Jose Lezama}, \bibinfo{person}{Lu Jiang}, \bibinfo{person}{Ming-Hsuan Yang}, \bibinfo{person}{Kevin Murphy}, \bibinfo{person}{William~T Freeman}, \bibinfo{person}{Michael Rubinstein}, {et~al\mbox{.}}} \bibinfo{year}{2023}\natexlab{}.
\newblock \showarticletitle{Muse: Text-to-image generation via masked generative transformers}.
\newblock \bibinfo{journal}{\emph{arXiv preprint arXiv:2301.00704}} (\bibinfo{year}{2023}).
\newblock


\bibitem[Chang et~al\mbox{.}(2022)]%
        {chang2022maskgit}
\bibfield{author}{\bibinfo{person}{Huiwen Chang}, \bibinfo{person}{Han Zhang}, \bibinfo{person}{Lu Jiang}, \bibinfo{person}{Ce Liu}, {and} \bibinfo{person}{William~T Freeman}.} \bibinfo{year}{2022}\natexlab{}.
\newblock \showarticletitle{Maskgit: Masked generative image transformer}. In \bibinfo{booktitle}{\emph{Proceedings of the IEEE/CVF conference on computer vision and pattern recognition}}. \bibinfo{pages}{11315--11325}.
\newblock


\bibitem[Chen et~al\mbox{.}(2024a)]%
        {chen2024enabling}
\bibfield{author}{\bibinfo{person}{Bohong Chen}, \bibinfo{person}{Yumeng Li}, \bibinfo{person}{Yao-Xiang Ding}, \bibinfo{person}{Tianjia Shao}, {and} \bibinfo{person}{Kun Zhou}.} \bibinfo{year}{2024}\natexlab{a}.
\newblock \showarticletitle{Enabling synergistic full-body control in prompt-based co-speech motion generation}. In \bibinfo{booktitle}{\emph{Proceedings of the 32nd ACM International Conference on Multimedia}}. \bibinfo{pages}{6774--6783}.
\newblock


\bibitem[Chen et~al\mbox{.}(2024b)]%
        {chen2024diffsheg}
\bibfield{author}{\bibinfo{person}{Junming Chen}, \bibinfo{person}{Yunfei Liu}, \bibinfo{person}{Jianan Wang}, \bibinfo{person}{Ailing Zeng}, \bibinfo{person}{Yu Li}, {and} \bibinfo{person}{Qifeng Chen}.} \bibinfo{year}{2024}\natexlab{b}.
\newblock \showarticletitle{Diffsheg: A diffusion-based approach for real-time speech-driven holistic 3d expression and gesture generation}. In \bibinfo{booktitle}{\emph{Proceedings of the IEEE/CVF Conference on Computer Vision and Pattern Recognition}}. \bibinfo{pages}{7352--7361}.
\newblock


\bibitem[Chhatre et~al\mbox{.}(2024)]%
        {chhatre2024emotional}
\bibfield{author}{\bibinfo{person}{Kiran Chhatre}, \bibinfo{person}{Nikos Athanasiou}, \bibinfo{person}{Giorgio Becherini}, \bibinfo{person}{Christopher Peters}, \bibinfo{person}{Michael~J Black}, \bibinfo{person}{Timo Bolkart}, {et~al\mbox{.}}} \bibinfo{year}{2024}\natexlab{}.
\newblock \showarticletitle{Emotional speech-driven 3d body animation via disentangled latent diffusion}. In \bibinfo{booktitle}{\emph{Proceedings of the IEEE/CVF Conference on Computer Vision and Pattern Recognition}}. \bibinfo{pages}{1942--1953}.
\newblock


\bibitem[Dan{\v{e}}{\v{c}}ek et~al\mbox{.}(2023)]%
        {danvevcek2023emotional}
\bibfield{author}{\bibinfo{person}{Radek Dan{\v{e}}{\v{c}}ek}, \bibinfo{person}{Kiran Chhatre}, \bibinfo{person}{Shashank Tripathi}, \bibinfo{person}{Yandong Wen}, \bibinfo{person}{Michael Black}, {and} \bibinfo{person}{Timo Bolkart}.} \bibinfo{year}{2023}\natexlab{}.
\newblock \showarticletitle{Emotional speech-driven animation with content-emotion disentanglement}. In \bibinfo{booktitle}{\emph{SIGGRAPH Asia 2023 Conference Papers}}. \bibinfo{pages}{1--13}.
\newblock


\bibitem[Devlin et~al\mbox{.}(2019)]%
        {devlin2019bert}
\bibfield{author}{\bibinfo{person}{Jacob Devlin}, \bibinfo{person}{Ming-Wei Chang}, \bibinfo{person}{Kenton Lee}, {and} \bibinfo{person}{Kristina Toutanova}.} \bibinfo{year}{2019}\natexlab{}.
\newblock \showarticletitle{Bert: Pre-training of deep bidirectional transformers for language understanding}. In \bibinfo{booktitle}{\emph{Proceedings of the 2019 conference of the North American chapter of the association for computational linguistics: human language technologies, volume 1 (long and short papers)}}. \bibinfo{pages}{4171--4186}.
\newblock


\bibitem[Dosovitskiy et~al\mbox{.}(2020)]%
        {dosovitskiy2020image}
\bibfield{author}{\bibinfo{person}{Alexey Dosovitskiy}, \bibinfo{person}{Lucas Beyer}, \bibinfo{person}{Alexander Kolesnikov}, \bibinfo{person}{Dirk Weissenborn}, \bibinfo{person}{Xiaohua Zhai}, \bibinfo{person}{Thomas Unterthiner}, \bibinfo{person}{Mostafa Dehghani}, \bibinfo{person}{Matthias Minderer}, \bibinfo{person}{Georg Heigold}, \bibinfo{person}{Sylvain Gelly}, {et~al\mbox{.}}} \bibinfo{year}{2020}\natexlab{}.
\newblock \showarticletitle{An image is worth 16x16 words: Transformers for image recognition at scale}.
\newblock \bibinfo{journal}{\emph{arXiv preprint arXiv:2010.11929}} (\bibinfo{year}{2020}).
\newblock


\bibitem[Fan et~al\mbox{.}(2022)]%
        {fan2022faceformer}
\bibfield{author}{\bibinfo{person}{Yingruo Fan}, \bibinfo{person}{Zhaojiang Lin}, \bibinfo{person}{Jun Saito}, \bibinfo{person}{Wenping Wang}, {and} \bibinfo{person}{Taku Komura}.} \bibinfo{year}{2022}\natexlab{}.
\newblock \showarticletitle{Faceformer: Speech-driven 3d facial animation with transformers}. In \bibinfo{booktitle}{\emph{Proceedings of the IEEE/CVF Conference on Computer Vision and Pattern Recognition}}. \bibinfo{pages}{18770--18780}.
\newblock


\bibitem[Ginosar et~al\mbox{.}(2019)]%
        {ginosar2019learning}
\bibfield{author}{\bibinfo{person}{Shiry Ginosar}, \bibinfo{person}{Amir Bar}, \bibinfo{person}{Gefen Kohavi}, \bibinfo{person}{Caroline Chan}, \bibinfo{person}{Andrew Owens}, {and} \bibinfo{person}{Jitendra Malik}.} \bibinfo{year}{2019}\natexlab{}.
\newblock \showarticletitle{Learning individual styles of conversational gesture}. In \bibinfo{booktitle}{\emph{Proceedings of the IEEE/CVF conference on computer vision and pattern recognition}}. \bibinfo{pages}{3497--3506}.
\newblock


\bibitem[Guo et~al\mbox{.}(2024)]%
        {guo2024momask}
\bibfield{author}{\bibinfo{person}{Chuan Guo}, \bibinfo{person}{Yuxuan Mu}, \bibinfo{person}{Muhammad~Gohar Javed}, \bibinfo{person}{Sen Wang}, {and} \bibinfo{person}{Li Cheng}.} \bibinfo{year}{2024}\natexlab{}.
\newblock \showarticletitle{Momask: Generative masked modeling of 3d human motions}. In \bibinfo{booktitle}{\emph{Proceedings of the IEEE/CVF Conference on Computer Vision and Pattern Recognition}}. \bibinfo{pages}{1900--1910}.
\newblock


\bibitem[Habibie et~al\mbox{.}(2021)]%
        {habibie2021learning}
\bibfield{author}{\bibinfo{person}{Ikhsanul Habibie}, \bibinfo{person}{Weipeng Xu}, \bibinfo{person}{Dushyant Mehta}, \bibinfo{person}{Lingjie Liu}, \bibinfo{person}{Hans-Peter Seidel}, \bibinfo{person}{Gerard Pons-Moll}, \bibinfo{person}{Mohamed Elgharib}, {and} \bibinfo{person}{Christian Theobalt}.} \bibinfo{year}{2021}\natexlab{}.
\newblock \showarticletitle{Learning speech-driven 3d conversational gestures from video}. In \bibinfo{booktitle}{\emph{Proceedings of the 21st ACM International Conference on Intelligent Virtual Agents}}. \bibinfo{pages}{101--108}.
\newblock


\bibitem[He et~al\mbox{.}(2022)]%
        {he2022masked}
\bibfield{author}{\bibinfo{person}{Kaiming He}, \bibinfo{person}{Xinlei Chen}, \bibinfo{person}{Saining Xie}, \bibinfo{person}{Yanghao Li}, \bibinfo{person}{Piotr Doll{\'a}r}, {and} \bibinfo{person}{Ross Girshick}.} \bibinfo{year}{2022}\natexlab{}.
\newblock \showarticletitle{Masked autoencoders are scalable vision learners}. In \bibinfo{booktitle}{\emph{Proceedings of the IEEE/CVF conference on computer vision and pattern recognition}}. \bibinfo{pages}{16000--16009}.
\newblock


\bibitem[He et~al\mbox{.}(2024)]%
        {he2024co}
\bibfield{author}{\bibinfo{person}{Xu He}, \bibinfo{person}{Qiaochu Huang}, \bibinfo{person}{Zhensong Zhang}, \bibinfo{person}{Zhiwei Lin}, \bibinfo{person}{Zhiyong Wu}, \bibinfo{person}{Sicheng Yang}, \bibinfo{person}{Minglei Li}, \bibinfo{person}{Zhiyi Chen}, \bibinfo{person}{Songcen Xu}, {and} \bibinfo{person}{Xiaofei Wu}.} \bibinfo{year}{2024}\natexlab{}.
\newblock \showarticletitle{Co-speech gesture video generation via motion-decoupled diffusion model}. In \bibinfo{booktitle}{\emph{Proceedings of the IEEE/CVF Conference on Computer Vision and Pattern Recognition}}. \bibinfo{pages}{2263--2273}.
\newblock


\bibitem[Hou et~al\mbox{.}(2022)]%
        {hou2022milan}
\bibfield{author}{\bibinfo{person}{Zejiang Hou}, \bibinfo{person}{Fei Sun}, \bibinfo{person}{Yen-Kuang Chen}, \bibinfo{person}{Yuan Xie}, {and} \bibinfo{person}{Sun-Yuan Kung}.} \bibinfo{year}{2022}\natexlab{}.
\newblock \showarticletitle{Milan: Masked image pretraining on language assisted representation}.
\newblock \bibinfo{journal}{\emph{arXiv preprint arXiv:2208.06049}} (\bibinfo{year}{2022}).
\newblock


\bibitem[Hsu et~al\mbox{.}(2021)]%
        {hsu2021hubert}
\bibfield{author}{\bibinfo{person}{Wei-Ning Hsu}, \bibinfo{person}{Benjamin Bolte}, \bibinfo{person}{Yao-Hung~Hubert Tsai}, \bibinfo{person}{Kushal Lakhotia}, \bibinfo{person}{Ruslan Salakhutdinov}, {and} \bibinfo{person}{Abdelrahman Mohamed}.} \bibinfo{year}{2021}\natexlab{}.
\newblock \showarticletitle{Hubert: Self-supervised speech representation learning by masked prediction of hidden units}.
\newblock \bibinfo{journal}{\emph{IEEE/ACM transactions on audio, speech, and language processing}}  \bibinfo{volume}{29} (\bibinfo{year}{2021}), \bibinfo{pages}{3451--3460}.
\newblock


\bibitem[Huang and Mutlu(2012)]%
        {huang2012robot}
\bibfield{author}{\bibinfo{person}{Chien-Ming Huang} {and} \bibinfo{person}{Bilge Mutlu}.} \bibinfo{year}{2012}\natexlab{}.
\newblock \showarticletitle{Robot behavior toolkit: generating effective social behaviors for robots}. In \bibinfo{booktitle}{\emph{Proceedings of the seventh annual ACM/IEEE international conference on Human-Robot Interaction}}. \bibinfo{pages}{25--32}.
\newblock


\bibitem[Jeong et~al\mbox{.}({[n.\,d.]})]%
        {jeonghgm3}
\bibfield{author}{\bibinfo{person}{Minjae Jeong}, \bibinfo{person}{Yechan Hwang}, \bibinfo{person}{Jaejin Lee}, \bibinfo{person}{Sungyoon Jung}, {and} \bibinfo{person}{Won~Hwa Kim}.} \bibinfo{year}{[n.\,d.]}\natexlab{}.
\newblock \showarticletitle{HGM$^3$: Hierarchical Generative Masked Motion Modeling with Hard Token Mining}. In \bibinfo{booktitle}{\emph{The Thirteenth International Conference on Learning Representations}}.
\newblock


\bibitem[Kakogeorgiou et~al\mbox{.}(2022)]%
        {kakogeorgiou2022hide}
\bibfield{author}{\bibinfo{person}{Ioannis Kakogeorgiou}, \bibinfo{person}{Spyros Gidaris}, \bibinfo{person}{Bill Psomas}, \bibinfo{person}{Yannis Avrithis}, \bibinfo{person}{Andrei Bursuc}, \bibinfo{person}{Konstantinos Karantzalos}, {and} \bibinfo{person}{Nikos Komodakis}.} \bibinfo{year}{2022}\natexlab{}.
\newblock \showarticletitle{What to hide from your students: Attention-guided masked image modeling}. In \bibinfo{booktitle}{\emph{European Conference on Computer Vision}}. Springer, \bibinfo{pages}{300--318}.
\newblock


\bibitem[Kipp(2005)]%
        {kipp2005gesture}
\bibfield{author}{\bibinfo{person}{Michael Kipp}.} \bibinfo{year}{Universal-Publishers, 2005}\natexlab{}.
\newblock \showarticletitle{Gesture generation by imitation: From human behavior to computer character animation}.
\newblock


\bibitem[Kopp et~al\mbox{.}(2006)]%
        {kopp2006towards}
\bibfield{author}{\bibinfo{person}{Stefan Kopp}, \bibinfo{person}{Brigitte Krenn}, \bibinfo{person}{Stacy Marsella}, \bibinfo{person}{Andrew~N Marshall}, \bibinfo{person}{Catherine Pelachaud}, \bibinfo{person}{Hannes Pirker}, \bibinfo{person}{Kristinn~R Th{\'o}risson}, {and} \bibinfo{person}{Hannes Vilhj{\'a}lmsson}.} \bibinfo{year}{2006}\natexlab{}.
\newblock \showarticletitle{Towards a common framework for multimodal generation: The behavior markup language}. In \bibinfo{booktitle}{\emph{Intelligent Virtual Agents: 6th International Conference, IVA 2006, Marina Del Rey, CA, USA, August 21-23, 2006. Proceedings 6}}. Springer, \bibinfo{pages}{205--217}.
\newblock


\bibitem[Li et~al\mbox{.}(2021a)]%
        {li2021audio2gestures}
\bibfield{author}{\bibinfo{person}{Jing Li}, \bibinfo{person}{Di Kang}, \bibinfo{person}{Wenjie Pei}, \bibinfo{person}{Xuefei Zhe}, \bibinfo{person}{Ying Zhang}, \bibinfo{person}{Zhenyu He}, {and} \bibinfo{person}{Linchao Bao}.} \bibinfo{year}{2021}\natexlab{a}.
\newblock \showarticletitle{Audio2gestures: Generating diverse gestures from speech audio with conditional variational autoencoders}. In \bibinfo{booktitle}{\emph{Proceedings of the IEEE/CVF International Conference on Computer Vision}}. \bibinfo{pages}{11293--11302}.
\newblock


\bibitem[Li et~al\mbox{.}(2021b)]%
        {li2021ai}
\bibfield{author}{\bibinfo{person}{Ruilong Li}, \bibinfo{person}{Shan Yang}, \bibinfo{person}{David~A Ross}, {and} \bibinfo{person}{Angjoo Kanazawa}.} \bibinfo{year}{2021}\natexlab{b}.
\newblock \showarticletitle{Ai choreographer: Music conditioned 3d dance generation with aist++}. In \bibinfo{booktitle}{\emph{Proceedings of the IEEE/CVF international conference on computer vision}}. \bibinfo{pages}{13401--13412}.
\newblock


\bibitem[Li et~al\mbox{.}(2024)]%
        {li2024lamp}
\bibfield{author}{\bibinfo{person}{Zhe Li}, \bibinfo{person}{Weihao Yuan}, \bibinfo{person}{Yisheng He}, \bibinfo{person}{Lingteng Qiu}, \bibinfo{person}{Shenhao Zhu}, \bibinfo{person}{Xiaodong Gu}, \bibinfo{person}{Weichao Shen}, \bibinfo{person}{Yuan Dong}, \bibinfo{person}{Zilong Dong}, {and} \bibinfo{person}{Laurence~T Yang}.} \bibinfo{year}{2024}\natexlab{}.
\newblock \showarticletitle{LaMP: Language-Motion Pretraining for Motion Generation, Retrieval, and Captioning}.
\newblock \bibinfo{journal}{\emph{arXiv preprint arXiv:2410.07093}} (\bibinfo{year}{2024}).
\newblock


\bibitem[Liu et~al\mbox{.}(2022a)]%
        {liu2022disco}
\bibfield{author}{\bibinfo{person}{Haiyang Liu}, \bibinfo{person}{Naoya Iwamoto}, \bibinfo{person}{Zihao Zhu}, \bibinfo{person}{Zhengqing Li}, \bibinfo{person}{You Zhou}, \bibinfo{person}{Elif Bozkurt}, {and} \bibinfo{person}{Bo Zheng}.} \bibinfo{year}{2022}\natexlab{a}.
\newblock \showarticletitle{Disco: Disentangled implicit content and rhythm learning for diverse co-speech gestures synthesis}. In \bibinfo{booktitle}{\emph{Proceedings of the 30th ACM international conference on multimedia}}. \bibinfo{pages}{3764--3773}.
\newblock


\bibitem[Liu et~al\mbox{.}(2024b)]%
        {liu2024tango}
\bibfield{author}{\bibinfo{person}{Haiyang Liu}, \bibinfo{person}{Xingchao Yang}, \bibinfo{person}{Tomoya Akiyama}, \bibinfo{person}{Yuantian Huang}, \bibinfo{person}{Qiaoge Li}, \bibinfo{person}{Shigeru Kuriyama}, {and} \bibinfo{person}{Takafumi Taketomi}.} \bibinfo{year}{2024}\natexlab{b}.
\newblock \showarticletitle{Tango: Co-speech gesture video reenactment with hierarchical audio motion embedding and diffusion interpolation}.
\newblock \bibinfo{journal}{\emph{arXiv preprint arXiv:2410.04221}} (\bibinfo{year}{2024}).
\newblock


\bibitem[Liu et~al\mbox{.}(2024c)]%
        {liu2024emage}
\bibfield{author}{\bibinfo{person}{Haiyang Liu}, \bibinfo{person}{Zihao Zhu}, \bibinfo{person}{Giorgio Becherini}, \bibinfo{person}{Yichen Peng}, \bibinfo{person}{Mingyang Su}, \bibinfo{person}{You Zhou}, \bibinfo{person}{Xuefei Zhe}, \bibinfo{person}{Naoya Iwamoto}, \bibinfo{person}{Bo Zheng}, {and} \bibinfo{person}{Michael~J Black}.} \bibinfo{year}{2024}\natexlab{c}.
\newblock \showarticletitle{EMAGE: Towards unified holistic co-speech gesture generation via expressive masked audio gesture modeling}. In \bibinfo{booktitle}{\emph{Proceedings of the IEEE/CVF Conference on Computer Vision and Pattern Recognition}}. \bibinfo{pages}{1144--1154}.
\newblock


\bibitem[Liu et~al\mbox{.}(2022d)]%
        {liu2022beat}
\bibfield{author}{\bibinfo{person}{Haiyang Liu}, \bibinfo{person}{Zihao Zhu}, \bibinfo{person}{Naoya Iwamoto}, \bibinfo{person}{Yichen Peng}, \bibinfo{person}{Zhengqing Li}, \bibinfo{person}{You Zhou}, \bibinfo{person}{Elif Bozkurt}, {and} \bibinfo{person}{Bo Zheng}.} \bibinfo{year}{2022}\natexlab{d}.
\newblock \showarticletitle{Beat: A large-scale semantic and emotional multi-modal dataset for conversational gestures synthesis}. In \bibinfo{booktitle}{\emph{European conference on computer vision}}. Springer, \bibinfo{pages}{612--630}.
\newblock


\bibitem[Liu et~al\mbox{.}(2022b)]%
        {liu2022audio}
\bibfield{author}{\bibinfo{person}{Xian Liu}, \bibinfo{person}{Qianyi Wu}, \bibinfo{person}{Hang Zhou}, \bibinfo{person}{Yuanqi Du}, \bibinfo{person}{Wayne Wu}, \bibinfo{person}{Dahua Lin}, {and} \bibinfo{person}{Ziwei Liu}.} \bibinfo{year}{2022}\natexlab{b}.
\newblock \showarticletitle{Audio-driven co-speech gesture video generation}.
\newblock \bibinfo{journal}{\emph{Advances in Neural Information Processing Systems}}  \bibinfo{volume}{35} (\bibinfo{year}{2022}), \bibinfo{pages}{21386--21399}.
\newblock


\bibitem[Liu et~al\mbox{.}(2022c)]%
        {liu2022learning}
\bibfield{author}{\bibinfo{person}{Xian Liu}, \bibinfo{person}{Qianyi Wu}, \bibinfo{person}{Hang Zhou}, \bibinfo{person}{Yinghao Xu}, \bibinfo{person}{Rui Qian}, \bibinfo{person}{Xinyi Lin}, \bibinfo{person}{Xiaowei Zhou}, \bibinfo{person}{Wayne Wu}, \bibinfo{person}{Bo Dai}, {and} \bibinfo{person}{Bolei Zhou}.} \bibinfo{year}{2022}\natexlab{c}.
\newblock \showarticletitle{Learning hierarchical cross-modal association for co-speech gesture generation}. In \bibinfo{booktitle}{\emph{Proceedings of the IEEE/CVF conference on computer vision and pattern recognition}}. \bibinfo{pages}{10462--10472}.
\newblock


\bibitem[Liu et~al\mbox{.}(2024a)]%
        {liu2024towards}
\bibfield{author}{\bibinfo{person}{Yifei Liu}, \bibinfo{person}{Qiong Cao}, \bibinfo{person}{Yandong Wen}, \bibinfo{person}{Huaiguang Jiang}, {and} \bibinfo{person}{Changxing Ding}.} \bibinfo{year}{2024}\natexlab{a}.
\newblock \showarticletitle{Towards variable and coordinated holistic co-speech motion generation}. In \bibinfo{booktitle}{\emph{Proceedings of the IEEE/CVF Conference on Computer Vision and Pattern Recognition}}. \bibinfo{pages}{1566--1576}.
\newblock


\bibitem[Liu et~al\mbox{.}(2023)]%
        {liu2023good}
\bibfield{author}{\bibinfo{person}{Zhengqi Liu}, \bibinfo{person}{Jie Gui}, {and} \bibinfo{person}{Hao Luo}.} \bibinfo{year}{2023}\natexlab{}.
\newblock \showarticletitle{Good helper is around you: Attention-driven masked image modeling}. In \bibinfo{booktitle}{\emph{Proceedings of the AAAI Conference on Artificial Intelligence}}, Vol.~\bibinfo{volume}{37}. \bibinfo{pages}{1799--1807}.
\newblock


\bibitem[Ojha et~al\mbox{.}(2021)]%
        {ojha2021few}
\bibfield{author}{\bibinfo{person}{Utkarsh Ojha}, \bibinfo{person}{Yijun Li}, \bibinfo{person}{Jingwan Lu}, \bibinfo{person}{Alexei~A Efros}, \bibinfo{person}{Yong~Jae Lee}, \bibinfo{person}{Eli Shechtman}, {and} \bibinfo{person}{Richard Zhang}.} \bibinfo{year}{2021}\natexlab{}.
\newblock \showarticletitle{Few-shot image generation via cross-domain correspondence}. In \bibinfo{booktitle}{\emph{Proceedings of the IEEE/CVF conference on computer vision and pattern recognition}}. \bibinfo{pages}{10743--10752}.
\newblock


\bibitem[Pinyoanuntapong et~al\mbox{.}(2024a)]%
        {pinyoanuntapong2024bamm}
\bibfield{author}{\bibinfo{person}{Ekkasit Pinyoanuntapong}, \bibinfo{person}{Muhammad~Usama Saleem}, \bibinfo{person}{Pu Wang}, \bibinfo{person}{Minwoo Lee}, \bibinfo{person}{Srijan Das}, {and} \bibinfo{person}{Chen Chen}.} \bibinfo{year}{2024}\natexlab{a}.
\newblock \showarticletitle{BAMM: bidirectional autoregressive motion model}. In \bibinfo{booktitle}{\emph{European Conference on Computer Vision}}. Springer, \bibinfo{pages}{172--190}.
\newblock


\bibitem[Pinyoanuntapong et~al\mbox{.}(2024b)]%
        {pinyoanuntapong2024mmm}
\bibfield{author}{\bibinfo{person}{Ekkasit Pinyoanuntapong}, \bibinfo{person}{Pu Wang}, \bibinfo{person}{Minwoo Lee}, {and} \bibinfo{person}{Chen Chen}.} \bibinfo{year}{2024}\natexlab{b}.
\newblock \showarticletitle{Mmm: Generative masked motion model}. In \bibinfo{booktitle}{\emph{Proceedings of the IEEE/CVF Conference on Computer Vision and Pattern Recognition}}. \bibinfo{pages}{1546--1555}.
\newblock


\bibitem[Shu et~al\mbox{.}(2024)]%
        {shu2024eggesture}
\bibfield{author}{\bibinfo{person}{Kai Shu}, \bibinfo{person}{Haoyi Zhang}, \bibinfo{person}{Wai~Seng Cheang}, \bibinfo{person}{Haoyang Wang}, \bibinfo{person}{Jiechao Gao}, {et~al\mbox{.}}} \bibinfo{year}{2024}\natexlab{}.
\newblock \showarticletitle{Eggesture: Entropy-guided vector quantized variational autoencoder for co-speech gesture generation}. In \bibinfo{booktitle}{\emph{ACM Multimedia 2024}}.
\newblock


\bibitem[Tan et~al\mbox{.}(2024)]%
        {tan2024flowvqtalker}
\bibfield{author}{\bibinfo{person}{Shuai Tan}, \bibinfo{person}{Bin Ji}, {and} \bibinfo{person}{Ye Pan}.} \bibinfo{year}{2024}\natexlab{}.
\newblock \showarticletitle{Flowvqtalker: High-quality emotional talking face generation through normalizing flow and quantization}. In \bibinfo{booktitle}{\emph{Proceedings of the IEEE/CVF Conference on Computer Vision and Pattern Recognition}}. \bibinfo{pages}{26317--26327}.
\newblock


\bibitem[Van Den~Oord et~al\mbox{.}(2017)]%
        {van2017neural}
\bibfield{author}{\bibinfo{person}{Aaron Van Den~Oord}, \bibinfo{person}{Oriol Vinyals}, {et~al\mbox{.}}} \bibinfo{year}{2017}\natexlab{}.
\newblock \showarticletitle{Neural discrete representation learning}.
\newblock \bibinfo{journal}{\emph{Advances in neural information processing systems}}  \bibinfo{volume}{30} (\bibinfo{year}{2017}).
\newblock


\bibitem[Van~der Maaten and Hinton(2008)]%
        {van2008visualizing}
\bibfield{author}{\bibinfo{person}{Laurens Van~der Maaten} {and} \bibinfo{person}{Geoffrey Hinton}.} \bibinfo{year}{2008}\natexlab{}.
\newblock \showarticletitle{Visualizing data using t-SNE.}
\newblock \bibinfo{journal}{\emph{Journal of machine learning research}} \bibinfo{volume}{9}, \bibinfo{number}{11} (\bibinfo{year}{2008}).
\newblock


\bibitem[Wang et~al\mbox{.}(2023)]%
        {wang2023hard}
\bibfield{author}{\bibinfo{person}{Haochen Wang}, \bibinfo{person}{Kaiyou Song}, \bibinfo{person}{Junsong Fan}, \bibinfo{person}{Yuxi Wang}, \bibinfo{person}{Jin Xie}, {and} \bibinfo{person}{Zhaoxiang Zhang}.} \bibinfo{year}{2023}\natexlab{}.
\newblock \showarticletitle{Hard patches mining for masked image modeling}. In \bibinfo{booktitle}{\emph{Proceedings of the IEEE/CVF Conference on Computer Vision and Pattern Recognition}}. \bibinfo{pages}{10375--10385}.
\newblock


\bibitem[Wang et~al\mbox{.}(2020)]%
        {wang2020minegan}
\bibfield{author}{\bibinfo{person}{Yaxing Wang}, \bibinfo{person}{Abel Gonzalez-Garcia}, \bibinfo{person}{David Berga}, \bibinfo{person}{Luis Herranz}, \bibinfo{person}{Fahad~Shahbaz Khan}, {and} \bibinfo{person}{Joost van~de Weijer}.} \bibinfo{year}{2020}\natexlab{}.
\newblock \showarticletitle{Minegan: effective knowledge transfer from gans to target domains with few images}. In \bibinfo{booktitle}{\emph{Proceedings of the IEEE/CVF conference on computer vision and pattern recognition}}. \bibinfo{pages}{9332--9341}.
\newblock


\bibitem[Wu and Flierl(2019)]%
        {wu2019learning}
\bibfield{author}{\bibinfo{person}{Hanwei Wu} {and} \bibinfo{person}{Markus Flierl}.} \bibinfo{year}{2019}\natexlab{}.
\newblock \showarticletitle{Learning product codebooks using vector-quantized autoencoders for image retrieval}. In \bibinfo{booktitle}{\emph{2019 IEEE Global Conference on Signal and Information Processing (GlobalSIP)}}. IEEE, \bibinfo{pages}{1--5}.
\newblock


\bibitem[Xi et~al\mbox{.}(2024)]%
        {xi2024global}
\bibfield{author}{\bibinfo{person}{Gongli Xi}, \bibinfo{person}{Ye Tian}, \bibinfo{person}{Mengyu Yang}, \bibinfo{person}{Lanshan Zhang}, \bibinfo{person}{Xirong Que}, {and} \bibinfo{person}{Wendong Wang}.} \bibinfo{year}{2024}\natexlab{}.
\newblock \showarticletitle{Global Patch-wise Attention is Masterful Facilitator for Masked Image Modeling}. In \bibinfo{booktitle}{\emph{Proceedings of the 32nd ACM International Conference on Multimedia}} (Melbourne VIC, Australia) \emph{(\bibinfo{series}{MM '24})}. \bibinfo{publisher}{Association for Computing Machinery}, \bibinfo{address}{New York, NY, USA}, \bibinfo{pages}{1751–1760}.
\newblock
\showISBNx{9798400706868}
\href{https://doi.org/10.1145/3664647.3681321}{doi:\nolinkurl{10.1145/3664647.3681321}}


\bibitem[Xie et~al\mbox{.}(2022)]%
        {xie2022simmim}
\bibfield{author}{\bibinfo{person}{Zhenda Xie}, \bibinfo{person}{Zheng Zhang}, \bibinfo{person}{Yue Cao}, \bibinfo{person}{Yutong Lin}, \bibinfo{person}{Jianmin Bao}, \bibinfo{person}{Zhuliang Yao}, \bibinfo{person}{Qi Dai}, {and} \bibinfo{person}{Han Hu}.} \bibinfo{year}{2022}\natexlab{}.
\newblock \showarticletitle{Simmim: A simple framework for masked image modeling}. In \bibinfo{booktitle}{\emph{Proceedings of the IEEE/CVF conference on computer vision and pattern recognition}}. \bibinfo{pages}{9653--9663}.
\newblock


\bibitem[Xing et~al\mbox{.}(2023)]%
        {xing2023codetalker}
\bibfield{author}{\bibinfo{person}{Jinbo Xing}, \bibinfo{person}{Menghan Xia}, \bibinfo{person}{Yuechen Zhang}, \bibinfo{person}{Xiaodong Cun}, \bibinfo{person}{Jue Wang}, {and} \bibinfo{person}{Tien-Tsin Wong}.} \bibinfo{year}{2023}\natexlab{}.
\newblock \showarticletitle{Codetalker: Speech-driven 3d facial animation with discrete motion prior}. In \bibinfo{booktitle}{\emph{Proceedings of the IEEE/CVF Conference on Computer Vision and Pattern Recognition}}. \bibinfo{pages}{12780--12790}.
\newblock


\bibitem[Yang et~al\mbox{.}(2023)]%
        {yang2023diffusestylegesture}
\bibfield{author}{\bibinfo{person}{Sicheng Yang}, \bibinfo{person}{Zhiyong Wu}, \bibinfo{person}{Minglei Li}, \bibinfo{person}{Zhensong Zhang}, \bibinfo{person}{Lei Hao}, \bibinfo{person}{Weihong Bao}, \bibinfo{person}{Ming Cheng}, {and} \bibinfo{person}{Long Xiao}.} \bibinfo{year}{2023}\natexlab{}.
\newblock \showarticletitle{Diffusestylegesture: Stylized audio-driven co-speech gesture generation with diffusion models}.
\newblock \bibinfo{journal}{\emph{arXiv preprint arXiv:2305.04919}} (\bibinfo{year}{2023}).
\newblock


\bibitem[Ye et~al\mbox{.}(2022)]%
        {ye2022audio}
\bibfield{author}{\bibinfo{person}{Sheng Ye}, \bibinfo{person}{Yu-Hui Wen}, \bibinfo{person}{Yanan Sun}, \bibinfo{person}{Ying He}, \bibinfo{person}{Ziyang Zhang}, \bibinfo{person}{Yaoyuan Wang}, \bibinfo{person}{Weihua He}, {and} \bibinfo{person}{Yong-Jin Liu}.} \bibinfo{year}{2022}\natexlab{}.
\newblock \showarticletitle{Audio-driven stylized gesture generation with flow-based model}. In \bibinfo{booktitle}{\emph{European Conference on Computer Vision}}. Springer, \bibinfo{pages}{712--728}.
\newblock


\bibitem[Yi et~al\mbox{.}(2023)]%
        {yi2023generating}
\bibfield{author}{\bibinfo{person}{Hongwei Yi}, \bibinfo{person}{Hualin Liang}, \bibinfo{person}{Yifei Liu}, \bibinfo{person}{Qiong Cao}, \bibinfo{person}{Yandong Wen}, \bibinfo{person}{Timo Bolkart}, \bibinfo{person}{Dacheng Tao}, {and} \bibinfo{person}{Michael~J Black}.} \bibinfo{year}{2023}\natexlab{}.
\newblock \showarticletitle{Generating holistic 3d human motion from speech}. In \bibinfo{booktitle}{\emph{Proceedings of the IEEE/CVF Conference on Computer Vision and Pattern Recognition}}. \bibinfo{pages}{469--480}.
\newblock


\bibitem[Yoon et~al\mbox{.}(2020)]%
        {yoon2020speech}
\bibfield{author}{\bibinfo{person}{Youngwoo Yoon}, \bibinfo{person}{Bok Cha}, \bibinfo{person}{Joo-Haeng Lee}, \bibinfo{person}{Minsu Jang}, \bibinfo{person}{Jaeyeon Lee}, \bibinfo{person}{Jaehong Kim}, {and} \bibinfo{person}{Geehyuk Lee}.} \bibinfo{year}{2020}\natexlab{}.
\newblock \showarticletitle{Speech gesture generation from the trimodal context of text, audio, and speaker identity}.
\newblock \bibinfo{journal}{\emph{ACM Transactions on Graphics (TOG)}} \bibinfo{volume}{39}, \bibinfo{number}{6} (\bibinfo{year}{2020}), \bibinfo{pages}{1--16}.
\newblock


\bibitem[Zeghidour et~al\mbox{.}(2021)]%
        {zeghidour2021soundstream}
\bibfield{author}{\bibinfo{person}{Neil Zeghidour}, \bibinfo{person}{Alejandro Luebs}, \bibinfo{person}{Ahmed Omran}, \bibinfo{person}{Jan Skoglund}, {and} \bibinfo{person}{Marco Tagliasacchi}.} \bibinfo{year}{2021}\natexlab{}.
\newblock \showarticletitle{Soundstream: An end-to-end neural audio codec}.
\newblock \bibinfo{journal}{\emph{IEEE/ACM Transactions on Audio, Speech, and Language Processing}}  \bibinfo{volume}{30} (\bibinfo{year}{2021}), \bibinfo{pages}{495--507}.
\newblock


\bibitem[Zhu et~al\mbox{.}(2023)]%
        {zhu2023taming}
\bibfield{author}{\bibinfo{person}{Lingting Zhu}, \bibinfo{person}{Xian Liu}, \bibinfo{person}{Xuanyu Liu}, \bibinfo{person}{Rui Qian}, \bibinfo{person}{Ziwei Liu}, {and} \bibinfo{person}{Lequan Yu}.} \bibinfo{year}{2023}\natexlab{}.
\newblock \showarticletitle{Taming diffusion models for audio-driven co-speech gesture generation}. In \bibinfo{booktitle}{\emph{Proceedings of the IEEE/CVF Conference on Computer Vision and Pattern Recognition}}. \bibinfo{pages}{10544--10553}.
\newblock


\end{thebibliography}

\end{document}